\documentclass[aps,prd,floats,floatfix,twoside,preprintnumbers,superscriptaddress,nofootinbib]{revtex4}

\usepackage{graphicx} 
\usepackage{dcolumn}
\usepackage{amssymb}
\usepackage{amsmath}
\usepackage[hidelinks]{hyperref}
\usepackage{color}
\usepackage{slashed}
\newcommand{\rmd}{\mathrm{d}}
\newcommand{\tr}{\mathrm{tr}}

\newcommand{\NC}{N_{\mathrm{c}}}

\newcommand{\MDEP}{\bp,\bq,\kg}

\newcommand{\bC}{\boldsymbol{C}}

\newcommand{\Uf}{\tilde{U}}
\newcommand{\vp}{v(\bp)}

\newcommand{\ubar}{\bar{u}(\bq)}

\newcommand{\rhop}{\rho_p}

\newcommand{\qbar}{\bar{q}}

\newcommand{\LIR}{\Lambda_{\rm IR}}

\newcommand{\calM}{\mathcal{M}}

\newcommand{\calN}{\mathcal{N}}

\newcommand{\xp}{\boldsymbol{x}_\perp} 
\newcommand{\yp}{\boldsymbol{y}_\perp} 

\newcommand{\Qp}{\boldsymbol{Q}_\perp} 
\newcommand{\Ppt}{\tilde{\boldsymbol{P}}_\perp} 

\newcommand{\khp}{\boldsymbol{k}_{1\perp}} 
\newcommand{\khps}{\slashed{\boldsymbol{k}}_{1\perp}} 
\newcommand{\kAp}{\boldsymbol{k}_{2\perp}} 
\newcommand{\kAps}{\slashed{\boldsymbol{k}}_{2\perp}} 
\newcommand{\kphp}{\boldsymbol{k}_{\gamma\perp}} 
\newcommand{\kg}{\boldsymbol{k}_\gamma} 
\newcommand{\kgp}{\boldsymbol{k}_{\gamma\perp}} 

\newcommand{\wperp}{\boldsymbol{w}_\perp}

\newcommand{\lp}{\boldsymbol{l}_\perp}

\newcommand{\kp}{\boldsymbol{k}_\perp}
\newcommand{\pp}{\boldsymbol{p}_\perp}
\newcommand{\Pp}{\boldsymbol{P}_\perp}
\newcommand{\qp}{\boldsymbol{q}_\perp}
\newcommand{\bk}{\boldsymbol{k}}
\newcommand{\bp}{\boldsymbol{p}}
\newcommand{\bq}{\boldsymbol{q}}
\newcommand{\kps}{\slashed{\bk}_\perp}

\newcommand{\vs}{\slashed{v}}
\newcommand{\ws}{\slashed{w}}

\newcommand{\Cs}{\slashed{C}}
\newcommand{\ks}{\slashed{k}}
\newcommand{\qs}{\slashed{q}}
\newcommand{\ps}{\slashed{p}}
\newcommand{\gs}{\slashed{g}}
\newcommand{\hs}{\slashed{h}}
\newcommand{\ls}{\slashed{l}}
\newcommand{\us}{\slashed{u}}

\newcommand{\msbar}{\overline{ \mathrm{MS}}} 

\begin{document}
\date{\today}

\title{Prompt photon - jet angular correlations at central rapidities in p+A collisions}
\author{Sanjin Beni\' c}
\affiliation{Physics Department, Faculty of Science,
                 University of Zagreb, Zagreb 10000, Croatia}
                 
                 \author{Adrian Dumitru}
\affiliation{Department of Natural Sciences, Baruch College, CUNY,
17 Lexington Avenue, New York, NY 10010, USA}
\affiliation{The Graduate School and University Center, The City University of New York,
365 Fifth Avenue, New York, NY 10016, USA}
\affiliation{Physics Department, Brookhaven National Lab, Upton, NY 11973, USA}
\begin{abstract} 
Photon-jet azimuthal correlations in proton-nucleus collisions are a
promising tool for gaining information on the gluon distribution of
the nucleus in the regime of non-linear color fields.  We compute such
correlations from the process $g\to q \bar{q} \gamma$ in the rapidity
regime where both the projectile and target light-cone momentum
fractions are small.  By integrating over the phase space of the quark
which emits the photon, subject to the restriction that the photon
picks up most of the transverse momentum (to pass an isolation cut),
we effectively obtain a $g+A\to q \gamma$ process.  For nearly
back-to-back photon-jet configurations we find that it dominates over
the leading order process $q+A\to q \gamma$ by two less powers of
$Q_\perp / Q_S$, where $\Qp$ and $Q_S$ denote the net photon-jet pair
momentum and the saturation scale of the nucleus, respectively. We
determine the transverse momentum dependent gluon distributions
involved in $g+A\to q \gamma$ and the scale where they are
evaluated. Finally, we provide analytic expressions for $\langle\cos
n\phi\rangle$ moments, where $\phi$ is the angle between $\Qp$ and the
average photon-jet transverse momentum $\Ppt$, and first qualitative
estimates of their transverse momentum dependence.
\end{abstract}


\maketitle 

\section{Introduction}

Angular correlations between hadrons and prompt photons in a
high-energy collisions have been suggested to provide insight into the
gluon fields of hadrons or nuclei in the small-$x$, non-linear
regime~\cite{JJM,Kopeliovich:1998nw,Gelis:2002ki,Baier:2004tj,Dominguez:2011br,Boer:2017xpy}.
Prompt photons are those originating from hard
  interactions among the initial beam partons.  Those studies have
mostly focussed on photon or dilepton production through the leading
order $q\to q\gamma$ process (in the field of the target) which
dominates in the fragmentation region of the projectile at forward
rapidities. In the central region on the other hand particle
production is dominated by processes involving soft partons from both
projectile and target. In particular, it has been pointed out in
ref.~\cite{Benic:2016uku} that the production of a photon can also
occur via $g\to q \bar{q} \gamma$. Although the amplitude squared for
this process is suppressed by one additional power of $\alpha_s$ as
compared to $q\to q\gamma$, on the other hand the gluon density in the
projectile at small $x$ formally is of order $1/\alpha_s$ times the
density of quarks and so these contributions should be comparable.

Here we consider angular correlations between photon with transverse
momentum $\kgp$ and a jet with transverse momentum $\pp$. We obtain
this final state by integrating over the phase space of the quark from
which the photon is emitted subject to the restriction that the photon
takes most of the transverse momentum. The contribution from the quark to photon fragmentation is explicitly taken into account. Hence, our process becomes
$g\to q\gamma$. We show that in the back-to-back correlation limit
where the photon-jet transverse momentum imbalance $\Qp=\pp + \kphp$
is much smaller than their average transverse momentum
$\Ppt=(\pp-\kphp)/2$, and smaller than the saturation scale $Q_S$ of
the nucleus, the $g\to q\gamma$ process in fact dominates
over $q\to q\gamma$ by two (fewer) powers of
$Q_\perp/Q_S$.

The $q\to q\gamma$ process can be written in terms of a convolution of
hard factors with the dipole forward scattering
amplitude~\cite{JJM,Kopeliovich:1998nw,Gelis:2002ki,Baier:2004tj,Boer:2017xpy}.
$g\to q \bar{q} \gamma$, on the other hand, in general involves the
expectation value of a correlator of four Wilson lines at small $x$
(see below and ref.~\cite{Benic:2016uku}).  In the near back-to-back
limit $Q_\perp^2/\tilde{P}_\perp^2\ll 1$, however, this correlator can
be expressed in terms of transverse momentum dependent gluon
distributions~\cite{Bomhof:2006dp,Dominguez:2011wm,Akcakaya:2012si}. Below, we
determine which gluon distributions appear in photon-jet production
through the $g\to q\gamma$ process, and at which transverse momentum
scale they are evaluated. This turns out to be given by the total
transverse momentum in the final state, thus the process can be used
to obtain information about the gluon distributions near the
non-linear ``saturation'' scale. Our focus here is on proton-nucleus
(p+A) collisions at high energies where the saturation scale $Q_S$ of the
heavy ion is expected to be semi-hard, on the order of a few GeV. We
employ the ``Color Glass Condensate''
formalism~\cite{KovchegovLevinBook} to describe the strong yet weakly
coupled gluon fields of the nucleus at small $x$.  For completeness
let us mention that photon-jet production in proton-proton (p+p) collisions
has been analyzed in the $k_T$-factorization approach with transverse
momentum dependent gluon distributions in
refs.~\cite{kT-fact}. Refs.~\cite{Klasen1,Klasen2} considered
photon-jet production in collinear factorization at NLO supplemented
by a parton shower generator. Lastly, such angular correlations from
$q\bar{q}\to \gamma g$ annihilation in $p\bar{p}$ collisions have been
discussed in ref.~\cite{Boer:2007nd}. Our main interest here is to
show how photon-jet azimuthal correlations at small-$x$ probe the
gluon fields of the heavy ion target in the non-linear regime.

From the cross section for $g\to q\gamma$ we can compute $\langle\cos
n\phi\rangle$ angular moments, where $\phi$ denotes the angle between
$\Qp$ and $\Ppt$. These are non-zero for all $n$ but increasingly
suppressed as $(Q_\perp/\tilde{P}_\perp)^n$. The transverse momentum
dependence of these moments in proton-nucleus collisions provides
information about the gluon distributions of the heavy ion target. We
also compute the ``azimuthal harmonics'' $\langle\cos n\Phi\rangle$
($n=1\dots3$) which are frequently studied in heavy-ion collisions,
where now $\Phi$ is the angle between $\kgp$ and $\pp$.

Photon-hadron (or jet) correlations have been studied experimentally
by the PHENIX~\cite{phenix} and STAR collaborations~\cite{star} at the
BNL-RHIC accelerator, and by CMS and ATLAS at the CERN-LHC. Thus far
the experiments at RHIC have focused on p+p or
nucleus-nucleus (A+A) collisions. The latter provide insight mainly into
final state interactions of the jet with the hot and dense medium
produced in heavy-ion collisions. Photon-jet correlations in p+p
collisions have provided information on transverse momentum dependent
gluon distributions albeit not in the weakly coupled, non-linear
regime of QCD since the saturation momentum of a proton at RHIC
energies (on average over impact parameters) is well below
1~GeV~\cite{AAMQS}.

The ATLAS collaboration has measured photon-jet correlations in p+p
collisions at 7~TeV~\cite{Aad:2013gaa}. For $p^{\rm jet}_\perp>40$~GeV and
$E_\perp^\gamma > 45$~GeV the correlations near $\phi=\pi$ can be
described reasonably well by established QCD Monte-Carlo generators. The
CMS collaboration, too, has analyzed these correlations in p+p and A+A
collisions but also obtained data for p+Pb collisions at
5~TeV~\cite{CMS:2013oua}.  Interestingly, for
$p_\perp^\mathrm{jet}>30$~GeV and 40~GeV $< k_\perp^\gamma<$ 50~GeV
the correlation strength near $\phi=\pi$ appears to be overestimated
by some Monte-Carlo models, see also Fig.~6 in
ref.~\cite{Klasen2}. This may potentially be related to high gluon
density effects of the kind considered here, especially if confirmed
by more symmetric configurations with smaller $Q_\perp$ (say, by going
to lower $k_\perp^\gamma> 30$~GeV).

This paper is organized as follows. In Sec.~\ref{sec:LO} we recall the
LO cross section expressed in terms of the transverse momentum
imbalance $\Qp$ and the average photon-jet average transverse momentum
$\Ppt$. The NLO amplitude is introduced in Sec.~\ref{sec:NLO}, and
expanded in powers of gluon momenta in the followup
Sec.~\ref{sec:expan}. In Sec.~\ref{sec:cs} we obtain the cross section
for $g \to q\qbar \gamma$, while in Sec.~\ref{sec:F_H_distributions} we
compute the nuclear distribution functions appearing in the NLO cross
section.  The $g \to q\gamma$ cross section and the angular moments
$a_n$ are obtained in Sec.~\ref{sec:CorrelLimit}. Our results are
summarized in Sec.~\ref{sec:summ}.

\section{Leading order photon-jet cross section}
\label{sec:LO}

\begin{figure}
  \begin{center}
  \includegraphics[scale = 0.5]{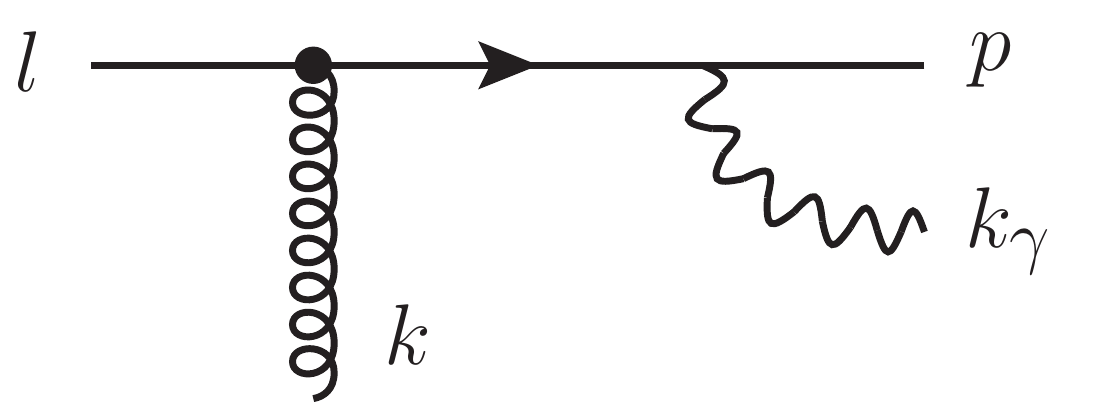}
  \end{center}
  \caption{Example for a LO diagram for photon production in the high-energy limit. The blob
    contains multiple (eikonal) scatterings on the dense target with
    momentum exchange $k$. The complete LO contribution includes
    photon emission before the interaction with the nucleus (not shown).}
  \label{fig:diaglo}
\end{figure}

The leading order (LO) photon-jet cross section due to scattering of a
(anti-) quark off the target is given by~\cite{Kopeliovich:1998nw,Gelis:2002ki,Baier:2004tj,Dominguez:2011br}
\begin{equation}
\begin{split}
&\frac{\rmd\sigma}{\rmd^2 \kgp \rmd \eta_{k_\gamma} \rmd^2 \pp \rmd \eta_p } = \frac{\alpha_e q_f^2}{32\pi^5}\int \rmd x_p [f_{q,p}(x_p,Q^2) + f_{\qbar,p}(x_p,Q^2)]\\
&\times\frac{1}{l^+}\left(l^{+2} + p^{+2}\right)\left[\frac{l\cdot p}{(l\cdot k_\gamma)(p\cdot k_\gamma)}+\frac{1}{p\cdot k_\gamma}-\frac{1}{l\cdot k_\gamma}\right]{\calN}_F(x_A,(\pp + \kgp)^2)(2\pi)\delta(l^+ - p^+ - k_{\gamma}^+)\,,
\end{split}
\end{equation}
where $k_\gamma$ and $p$ are the photon and jet momenta, respectively,
while $l$ is the momentum of the quark from the proton, see Fig.~\ref{fig:diaglo}. Here
$f_{q(\qbar),p}(x_p,Q^2)$ denote the collinear quark and the antiquark
distribution functions of the proton evaluated at some hard scale
$Q^2$ and ${\calN}_F(x_A,k_\perp^2)$ is the forward scattering
amplitude of a fundamental dipole off the nucleus. At $x_A\ll 1$ and
in the large-$N_c$ limit this is obtained by solving the
Balitsky-Kovchegov (BK) evolution equation~\cite{BK}.

We can rewrite the cross section in terms of the light cone momentum
fraction of the outgoing quark $z_q=p^+/l^+$ and the transverse momentum
imbalance $\Qp$ and the average transverse momentum $\Ppt$ which are
defined via
\begin{equation}
  \Qp \equiv \kgp + \pp \,, \qquad
\Ppt \equiv \frac{1}{2}(\pp - \kgp)\,.  
\label{eq:imb}
\end{equation}

In terms of these variables,
\begin{equation}
\begin{split}
&\frac{\rmd\sigma}{\rmd^2 \Ppt \rmd^2 \Qp \rmd
    \eta_{k_\gamma} \rmd \eta_p } = \frac{\alpha_e q_f^2}{16\pi^5}\int
  \rmd x_p [f_{q,f}(x_p,Q^2) + f_{\qbar,f}(x_p,Q^2)]\\
  &\times
  \frac{1 + z_q^2}{z_q}\frac{z_q (1-z_q)^2
    Q_\perp^2}{\left(\frac{1}{2}\Qp - \Ppt\right)^2
    \left[\left(\frac{1}{2}-z_q\right)\Qp +
      \Ppt\right]^2} \, {\calN}_F(x_A,Q_\perp^2)\,
  (2\pi)p^+ \delta(l^+ - p^+ - k_{\gamma}^+)\,.
\end{split}
\end{equation}
We focus on photon-jet configurations that correspond to
$\tilde{P}_\perp \gg Q_S\gg Q_\perp$ so that the transverse momenta of
the photon and the jet are almost back to back. In this limit, due to
non-linear effects the dipole forward scattering amplitude
$\calN_F(x_A,Q_\perp^2)\sim 1/Q_S^2(x_A)$ is proportional to the inverse
saturation scale squared of the nucleus. Hence, the leading
contribution to the LO cross section is of order $(Q_\perp/Q_S)^2$,
times an overall $1/\tilde{P}_\perp^4$, and is independent of the
angle $\phi$ between $\Qp$ and $\Ppt$. An angular dependence $\sim
\cos \phi$ arises at next-to leading power $(Q_\perp/Q_S)^2\,
\Qp\cdot\Ppt/\tilde{P}_\perp^2$.  Below we show that the cross section
for $g\to q\gamma$ generates an isotropic contribution at order 1
(times the common $1/\tilde{P}_\perp^4$), a $\cos \phi$ angular
dependence already at next-to-leading power $\sim
Q_\perp/\tilde{P}_\perp$, and so on.

\section{Next to leading order photon-jet amplitude}
\label{sec:NLO}

\begin{figure}
  \begin{center}
  \includegraphics[scale = 0.5]{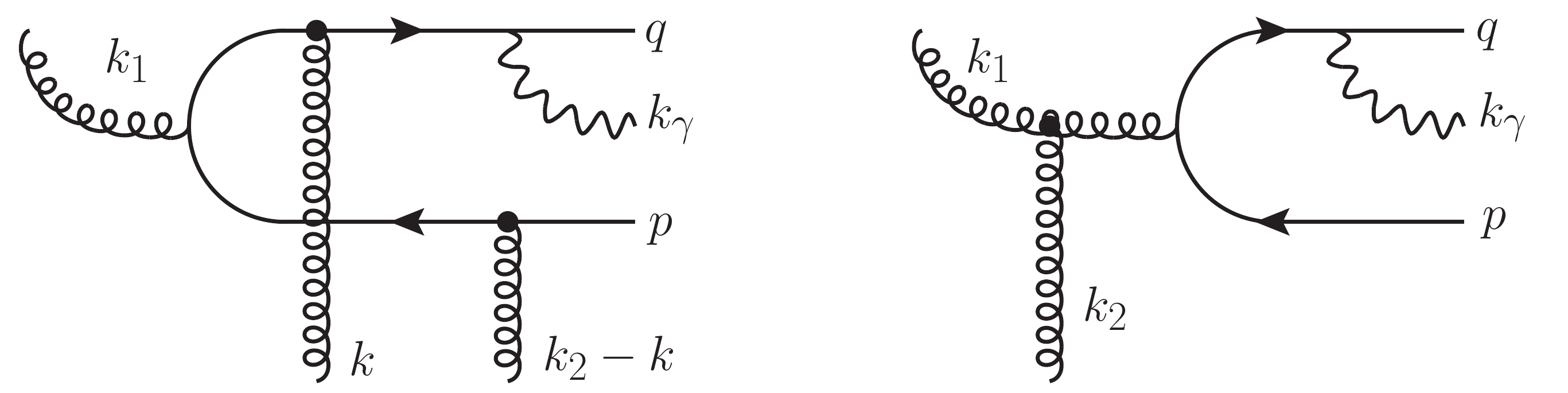}
  \end{center}
  \caption{Examples for NLO diagrams for photon production in the high-energy limit. In the
    left diagram the gluon from the proton emits a $q\bar{q}$ pair
    which then scatters on the dense nucleus. In the right diagram the
    gluon scatters on the nucleus and then produces a $q\bar{q}$ pair
    and a photon. The complete NLO contribution includes
    diagrams where the photon is emitted from the antiquark line, or from
    a virtual quark anti-quark state (not shown) in case of
    single-inclusive photon production.}
  \label{fig:diagnlo}
\end{figure}

At the NLO order photons are produced via the $g\to q \bar{q}
\gamma$ process where either the incoming gluon or either of the
quarks may scatter off the field of the target, see
Fig.~\ref{fig:diagnlo}. For a photon-jet final state, one of the
quarks will eventually be integrated over\footnote{Specifically, we
  integrate over the phase space of the quark which emits the
  photon. We restrict to the region of phase space where the photon
  picks up most of the momentum of the parent quark so that the
  configuration passes a photon isolation cut, see Sec.~\ref{sec:jetgamma} below}.. The external momenta
in the process are $\boldsymbol{q}$, $\boldsymbol{p}$ and $\boldsymbol{k}_\gamma$ for the quark, antiquark
and the photon, respectively.  In the amplitude we also have the transverse gluon
momenta from the proton, $\khp$, and from the nucleus, $\kp$ and $\kAp
- \kp$ (there are two, because of the two Wilson lines). Here, $\kAp
\equiv \Pp - \khp$ with $\Pp \equiv \qp + \pp + \kgp$ is the total
final state transverse momentum.

The main formula for the amplitude is (see Eq.~(47) in \cite{Benic:2016uku})
\begin{equation}
  \begin{split}
    \calM^\mu(\MDEP) &= -q_f e g^2\int_{\kp \khp}\int_{\xp\yp}
    \frac{\rhop^a(\khp)}{k_{1\perp}^2}\,
    e^{i\kp\cdot\xp+i(\kAp-\kp)\cdot\yp} \\
    &\qquad\times \ubar \bigl\{ T_g^\mu(\khp)U(\xp)^{ba}t^b
    +T_{q\qbar}^\mu(\kp,\khp)\Uf(\xp)t^a\Uf^\dag(\yp)\bigr\}\vp\,.
  \end{split}
\label{eq:full-amp}
\end{equation}
Here $\tilde{U}(\xp)$ ($U(\xp)$) is the Wilson line in the fundamental (adjoint) representation
\begin{equation}
\begin{split}
\tilde{U}(\xp) \equiv \mathcal{P}_+ \exp\left[-ig^2\int_{-\infty}^\infty dz^+ \frac{1}{\boldsymbol{\nabla}_\perp^2}\rho^a_A(z^+,\xp) t^a\right]\,,\\
U(\xp) \equiv \mathcal{P}_+ \exp\left[-ig^2\int_{-\infty}^\infty dz^+ \frac{1}{\boldsymbol{\nabla}_\perp^2}\rho^a_A(z^+,\xp) T^a\right]\,,\\
\end{split}
\end{equation}
with $t^a$ ($T^a$) the $SU(N_c)$ generators in the fundamental
(adjoint) representation. $g\rho^a_p(\xp)$ ($g\rho^a_A(\xp)$) denotes
the random color charge density of the proton (nucleus) which will be
averaged over after squaring the amplitude. The other factor of the
coupling $g$ in Eq.~(\ref{eq:full-amp}) arises from the $g-q-\bar{q}$ vertex.

We also introduce the following momentum labels
\begin{equation}
w \equiv q+k_\gamma \,, \qquad v \equiv p+k_\gamma \,, \qquad g \equiv q-k \,, \qquad h \equiv q-k-k_1 \,, \qquad u \equiv q+k_\gamma - k \,, \qquad l \equiv q+k_\gamma - k - k_1 \,,
\label{eq:mom}
\end{equation}
where $k^+ = 0$, $k_1^- = 0$, $k^+_1 = P^+$.
We also define the Lipatov vertex
\begin{equation}
  C_L^+(q,\khp) = q^+-\frac{k_{1\perp}^2}{q^-+i\epsilon}\,, \quad
  C_L^-(q,\khp) = \frac{(\qp-\khp)^2}{q^+}-q^-\,, \quad
  \bC_{L\perp}(q,\khp) = \qp - 2\khp\,.
\label{eq:Lip2}
\end{equation}
We then have
\begin{equation}
T^\mu_g \equiv \sum_{\beta=1}^2 R_\beta^\mu \,, \qquad 
T^\mu_{q\qbar} \equiv \sum_{\beta=9}^{12} R_\beta^\mu \,,
\end{equation}
with
\begin{equation}
R_1^\mu = -\frac{\gamma^\mu(\ws + m)\Cs_L}{P^2 (w^2 - m^2)}\,, \qquad R_2^\mu = -\frac{\Cs_L(-\vs + m)\gamma^\mu}{P^2 (v^2 - m^2)}\,,
\label{eq:R1R2mu}
\end{equation}
\begin{equation}
R_9^\mu = -\frac{\gamma^\mu(\ws + m)\gamma^+ (\us + m)\gamma^- (\ls + m)\gamma^+}{N_k (w^2 - m^2)}\,, \qquad N_k = 2p^+ M^2(u_\perp) + 2 w^+ M^2(l_\perp) \,,
\label{eq:R9mu}
\end{equation}
\begin{equation}
R_{10}^\mu = -\frac{\gamma^+(\gs + m)\gamma^- (\hs + m)\gamma^+ (- \vs + m)\gamma^\mu}{N_q (v^2 - m^2)}\,, \qquad N_q \equiv 2v^+ M^2(g_\perp) + 2 q^+ M^2(h_\perp) \,,
\label{eq:R10mu}
\end{equation}
\begin{equation}
R_{11}^\mu = 2p^+\frac{\gamma^+(\gs + m)\gamma^\mu (\us + m)\gamma^- (\ls + m)\gamma^+}{N_k S}\,, \qquad k^- = w^-  + \frac{M^2(\lp)}{2p^+}\,,
\label{eq:R11mu}
\end{equation}
\begin{equation}
R_{12}^\mu = 2q^+\frac{\gamma^+(\gs + m)\gamma^- (\hs + m)\gamma^\mu (\ls + m)\gamma^+}{N_q S}\,, \qquad k^- = q^- - \frac{M^2(g_\perp)}{2 q^+}\,,
\label{eq:R12mu}
\end{equation}
with
$$S = 4 p^+ q^+ k_\gamma^- + 2q^+ M^2(l_\perp) + 2p^+ M^2(g_\perp)\,.$$
Here we defined the transverse mass function $M(k_\perp) \equiv \sqrt{k_\perp^2 + m^2}$.

We assume that there is a certain combination of the external momenta
which is the largest scale in the process (i.e.\ the average
transverse momentum $\Ppt$ of photon and jet). This scale is also
supposed to be much larger than the saturation scale of the proton or
that of the nucleus. The transverse momenta of the gluon from the
proton, $k_{1\perp}$, and of the gluons from the nucleus, $k_\perp$
and $|\kAp - \kp|$, are of the order of the respective saturation
scales, much smaller than the hard scale.

With this assumption we perform an expansion of the amplitude in
$\khp$, $\kAp - \kp$ and $\kp$. To obtain the gluon distributions we
must expand to first order in these momenta.  The main point is that
by Ward identities this expansion necessarily involves terms
proportional to $k_{1\perp}^i (k_2 - k_\perp)^j$ or $k_{1\perp}^i
k_\perp^j$ in the amplitude. Note that by conservation of momentum
this implies that the total transverse momentum $\Pp \equiv \qp + \pp
+ \kgp$ in the final state is much smaller than the average transverse
momentum $\Ppt$ of the photon and jet. We shall return to this point in Sec.~\ref{sec:CorrelLimit}.
After some algebra, the full details are given in
  App.~\ref{sec:expan}, we obtain the following main result for the amplitude:
\begin{equation}
  \begin{split}
    \calM^\mu(\MDEP) &= -q_f e g^2\int_{\kp \khp}\int_{\xp\yp}
    \frac{\rhop^a(\khp)}{k_{1\perp}^2}\,
    e^{i\kp\cdot\xp+i(\Pp-\kp-\khp)\cdot\yp} \\
    &\qquad\times \ubar \left\{ k_{2 i} R_g^{\mu i}U(\xp)^{ba}t^b
    + \left[k_i R^{\mu i}_q + (k_2 - k)_i R^{\mu i}_{\bar{q}}\right]\Uf(\xp)t^a\Uf^\dag(\yp)\right\}\vp\,,
  \end{split}
\label{eq:amp3}
\end{equation}
where we defined
\begin{equation}
R^{\mu i}_g \equiv \sum_{\beta = 1}^2 R^{\mu i}_\beta\,,\qquad R^{\mu i}_q \equiv \sum_{\beta = 3}^5 R^{\mu i}_\beta\,,\qquad R^{\mu i}_{\qbar} \equiv \sum_{\beta = 6}^8 R^{\mu i}_\beta\,,
\label{eq:Rs}
\end{equation}
with
\begin{equation}
\begin{split}
& R^{\mu i}_1 \equiv \frac{2}{P^2}\gamma^\mu \frac{\qs + \ks_\gamma + m}{(q+k_\gamma)^2 - m^2}\ks_2\frac{k_1^i}{P^+ P^-}\,,\\
& R^{\mu i}_2 \equiv \frac{2}{P^2}\ks_2 \frac{-\ps - \ks_\gamma + m}{(p+k_\gamma)^2 - m^2}\gamma^\mu\frac{k_1^i}{P^+ P^-}\,,\\
& R^{\mu i}_3 \equiv - \gamma^\mu \frac{\qs + \ks_\gamma + m}{(q+k_\gamma)^2 - m^2} \frac{\gamma^i}{P^-}\frac{\ks_1 - \ps + m}{(k_1 - p)^2 - m^2}\frac{\khps}{P^+}\,,\\
& R^{\mu i}_4 \equiv - \frac{\gamma^i}{P^-}\frac{\qs - \ks_2 + m}{(q-k_2)^2 - m^2}\gamma^\mu \frac{\ks_1 - \ps + m}{(k_1 - p)^2 - m^2}\frac{\khps}{P^+}\,,\\
& R^{\mu i}_5 \equiv - \frac{\gamma^i}{P^-} \frac{\qs - \ks_2 + m}{(q-k_2)^2 - m^2} \frac{\khps}{P^+}  \frac{-\ps - \ks_\gamma + m}{(p + k_\gamma)^2 - m^2}\gamma^\mu\,,\\
& R^{\mu i}_6 \equiv \frac{\khps}{P^+} \frac{\qs - \ks_1 + m}{(q-k_1)^2 - m^2}\frac{\gamma^i}{P^-}\frac{-\ps - \ks_\gamma + m}{(p+k_\gamma)^2 - m^2}\gamma^\mu \,,\\
& R^{\mu i}_7 \equiv \frac{\khps}{P^+} \frac{\qs - \ks_1 + m}{(q-k_1)^2 - m^2}\gamma^\mu \frac{\ks_2 - \ps + m}{(k_2 - p)^2 - m^2}\frac{\gamma^i}{P^-}\,,\\
& R^{\mu i}_8 \equiv \gamma^\mu \frac{\qs + \ks_\gamma + m}{(q+k_\gamma)^2 - m^2}\frac{\khps}{P^+} \frac{\ks_2 - \ps + m}{(k_2 - p)^2 - m^2}\frac{\gamma^i}{P^-}\,.\\
\end{split}
\label{eq:har1}
\end{equation}
Here $R_\beta^{\mu i}$, with $\beta = 3,\dots 5$ ($\beta = 6,\dots,
8$) correspond to the three terms in the first (second) line of
Eq.~\eqref{eq:Rqqb}, after using the momentum definitions
Eq.~\eqref{eq:mom} and the collinear limit:
\begin{equation}
\begin{split}
&k_1 = (k_1^+,k_1^-,\khp) = (P^+,0,0)\,,\\
&k_2 = (k_2^+,k_2^-,\kAp) = (0,P^-,0)\,.
\end{split}
\end{equation}
As an aside, it is useful to check the soft photon limit of the
amplitude in Eq.~\eqref{eq:amp3}. We obtain
\begin{equation}
  \begin{split}
    \calM^\mu(\MDEP) &= q_f e \left(\frac{p^\mu}{p\cdot k_\gamma}-\frac{q^\mu}{q\cdot k_\gamma}\right) g^2\int_{\kp \khp}\int_{\xp\yp}
    \frac{\rhop^a(\khp)}{k_{1\perp}^2}\,
    e^{i\kp\cdot\xp+i(\Pp-\kp-\khp)\cdot\yp} \\
    &\qquad\times \ubar \left\{ k_{2 i} R_g^i U(\xp)^{ba}t^b
    + \left[k_i R^i_q + (k_2 - k)_i R^i_{\bar{q}}\right]\Uf(\xp)t^a\Uf^\dag(\yp)\right\}\vp\,,
  \end{split}
\label{eq:ampsoft}
\end{equation}
where
\begin{equation}
\begin{split}
& R_g^i \equiv -\frac{2}{P^2}\ks_2\frac{k_1^i}{P^+ P^-}\,,\\
& R_q^i \equiv \frac{\gamma^i}{P^-}\frac{\ks_1 - \ps + m}{(k_1 - p)^2 - m^2}\frac{\khps}{P^+}\,,\\
& R^i_{\qbar} \equiv -\frac{\khps}{P^+} \frac{\ks_2 - \ps + m}{(k_2 - p)^2 - m^2}\frac{\gamma^i}{P^-}\,.\\
\end{split}
\label{eq:Rsoft}
\end{equation}
Thus, Eq.~\eqref{eq:ampsoft} factorizes into the soft photon emission
amplitude times the amplitude for $q\qbar$ production. The
$q\qbar$ production part exactly agrees with the result in
Ref.~\cite{Akcakaya:2012si} once the expressions in Eqs. (25) and (26)
of \cite{Akcakaya:2012si} are compared with Eq.~\eqref{eq:Rsoft}
through $R_{g,i} = \tilde{T}_{g,i}$, $R_{q,i} =
\tilde{T}_{q\qbar,i}^A$ and $R_{\qbar,i} = \tilde{T}_{q\qbar,i}^B$
together with $k_1 \to k_2$ and $k_2 \to k_1$ in the notation of
Ref.~\cite{Akcakaya:2012si}.

\section{Next to leading order cross section}
\label{sec:cs}

We can immediately derive the cross section by adapting the
calculation from ref.~\cite{Benic:2016uku}. With the replacements
\begin{equation}
k_{2i}R^{\mu i}_g \to T^\mu_g \,, \qquad k_i R^{\mu i}_q + (k_2 - k)_i R^{\mu i}_{\qbar} \to T^\mu_{q\qbar} \,.
\end{equation}
the amplitude in Eq.~\eqref{eq:amp3} has the same structure as \eqref{eq:full-amp} (or Eq.~(47) in \cite{Benic:2016uku}). Then, we can deduce the cross section from Eq.~(63) in \cite{Benic:2016uku}
\begin{equation}
\begin{split}
&\frac{d\sigma}{d^2 \pp d^2 \qp d^2 \kgp d\eta_p d\eta_q d\eta_{k_\gamma}} = \frac{\alpha_e\alpha_S q_f^2}{256 \pi^8 N_c(N_c^2 - 1)}\int_{\khp \kAp}(2\pi)^2 \delta^{(2)}(\Pp - \khp - \kAp) \frac{\varphi_p(x_p,k_{1\perp}^2)}{k_{1\perp}^2}\\
&\times \int_{\xp\xp'\yp\yp'}\int_{\kp,\kp'} e^{i(\kp\cdot\xp - \kp'\cdot\xp')+i(\kAp - \kp)\cdot\yp - i(\kAp - \kp')\cdot\yp'} g^{\mu\mu'}\\
&\times\Bigl\{\tr\left[(\qs + m)k_2^i R_{g,\mu i}(m-\ps)\gamma^0 k_2^{i'} R'^\dag_{g,\mu' i'}\gamma^0\right]C(x_A,\xp,\xp,\xp',\xp')\\
& + \tr\left[(\qs + m)k_2^i R_{g,\mu i}(m-\ps)\gamma^0 \left(k'^{i'} R'^{\dag}_{q,\mu'i'} + (k_2 - k')^{i'} R'^\dag_{\qbar,\mu' i'}\right)\gamma^0\right]C(x_A,\xp,\xp,\yp',\xp')\\
& + \tr\left[(\qs + m)\left(k^i R_{q,\mu i} + (k_2 - k)^i R_{\qbar,\mu i}\right)(m-\ps)\gamma^0 k_2^{i'} R'^\dag_{g,\mu' i'}\gamma^0\right]C(x_A,\xp,\yp,\xp',\xp')\\
& + \tr\left[(\qs + m)\left(k^i R_{q,\mu i} + (k_2 - k)^i R_{\qbar,\mu i}\right)(m-\ps)\gamma^0 \left(k'^{i'} R'^{\dag}_{q,\mu'i'} + (k_2 - k')^{i'} R'^\dag_{\qbar,\mu' i'}\right)\gamma^0\right]C(x_A,\xp,\yp,\yp',\xp')\Bigr\}\,.
\end{split}
\end{equation}
To follow more closely the notation in \cite{Akcakaya:2012si} we
rewrote the nuclear gluon field correlators from Eqs.~(58)--(60) in
ref.~\cite{Benic:2016uku} in terms of the four Wilson line correlator
\begin{equation}
C(x_A,\xp,\yp,\yp',\xp') \equiv \tr_c\left\langle \tilde{U}(\xp)t^a \tilde{U}^\dag(\yp)\tilde{U}(\yp') t^a \tilde{U}^\dag(\xp')\right\rangle_{x_A}\,.
\label{eq:corrnuc}
\end{equation}
We used $t^b U^{ba}(\xp) = \tilde{U}(\xp)t^a \tilde{U}(\xp)$. Also, we
defined the unintegrated gluon distribution
$\varphi_p(x_p,k_{1\perp}^2)$ of the proton via
\begin{equation}
  \left\langle\rhop^a(\khp)\rhop^{\dag b}(\khp)\right\rangle_{x_p}
  \equiv \frac{\delta^{ab}k_{1\perp}^2}{2\pi \NC C_F\, g^2}\,\varphi_p(x_p,k_{1\perp}^2)\,.
\label{eq:rho_corrproton}
\end{equation}
The proton and the nucleus light-cone momentum fractions, $x_p$ and $x_A$,
are related to the final state kinematics via
\begin{equation}
\begin{split}
& x_p \sqrt{\frac{s}{2}} = p^+ + k_\gamma^+ + q^+\,,\\
& x_A \sqrt{\frac{s}{2}} = p^- + k_\gamma^- + q^-\,.
\end{split}
\end{equation}

By partial integration we can turn powers of $k^i$, $(k_2
- k)^i$ and $k_2^i$ into gradients of the correlator of four Wilson lines \eqref{eq:corrnuc}.
For example,
\begin{equation}
\begin{split}
&\int_{\xp\xp'\yp\yp'}\int_{\kp,\kp'} e^{i(\kp\cdot\xp - \kp'\cdot\xp')+i(\kAp - \kp)\cdot\yp - i(\kAp - \kp')\cdot\yp'} k_2^i (k_2 - k')^{i'}C(x_A,\xp,\xp,\yp',\xp')\\
& = \int_{\xp\xp'\yp'}\int_{\kp'} e^{-i\kp'\xp' + i\kAp\xp - i(\kAp - \kp')\cdot\yp'}k_2^i (k_2 - k')^{i'}C(x_A,\xp,\xp,\yp',\xp')\\
& = \int_{\xp\xp'\yp'}\int_{\kp'} e^{-i\kp'\xp' + i\kAp\xp - i(\kAp - \kp')\cdot\yp'} \frac{\partial^2 C(x_A,\xp,\xp,\yp',\xp')}{\partial x^i \partial y'^{i'}}\\
&= \int_{\xp\xp'} e^{i\kAp\cdot (\xp -\xp')} \left[\frac{\partial^2 C(x_A,\xp,\xp,\yp',\xp')}{\partial x^i \partial y'^{i'}}\right]_{\xp' = \yp'}\,.
\end{split}
\end{equation}
The cross section can be written as
\begin{equation}
\begin{split}
&\frac{d\sigma}{d^2 \pp d^2 \qp d^2 \kgp d\eta_p d\eta_q d\eta_{k_\gamma}} = \frac{\alpha_e\alpha_S q_f^2}{256 \pi^8 N_c(N_c^2 - 1)}\int_{\khp \kAp}(2\pi)^2 \delta^{(2)}(\Pp - \khp - \kAp) \frac{\varphi_p(x_p,k_{1\perp}^2)}{k_{1\perp}^2}\\
&\times\int_{\xp \xp'} e^{i\kAp\cdot(\xp - \xp')} g^{\mu\mu'}\\
\times\Biggl\{&\tr\left[(\qs + m)R_{q,\mu i}(m-\ps)\gamma^0 R'^\dag_{q,\mu' i'}\gamma^0\right]\left[\frac{\partial^2 C(x_A,\xp,\yp,\yp',\xp')}{\partial x^i \partial x'^{i'}}\right]_{\xp = \yp, \xp' =  \yp'}\\
+&\tr\left[(\qs + m)R_{q,\mu i}(m-\ps)\gamma^0 R'^\dag_{\bar{q},\mu' i'}\gamma^0\right]\left[\frac{\partial^2 C(x_A,\xp,\yp,\yp',\xp')}{\partial x^i \partial y'^{i'}}\right]_{\xp = \yp, \xp' =  \yp'}\\
+&\tr\left[(\qs + m)R_{\bar{q},\mu i}(m-\ps)\gamma^0 R'^\dag_{q,\mu' i'}\gamma^0\right]\left[\frac{\partial^2 C(x_A,\xp,\yp,\yp',\xp')}{\partial y^i \partial x'^{i'}}\right]_{\xp = \yp, \xp' =  \yp'}\\
+&\tr\left[(\qs + m)R_{\bar{q},\mu i}(m-\ps)\gamma^0 R'^\dag_{\bar{q},\mu' i'}\gamma^0\right]\left[\frac{\partial^2 C(x_A,\xp,\yp,\yp',\xp')}{\partial y^i \partial y'^{i'}}\right]_{\xp = \yp, \xp' =  \yp'}\\
+&\tr\left[(\qs + m)R_{q,\mu i}(m-\ps)\gamma^0 R'^\dag_{g,\mu' i'}\gamma^0\right]\left[\frac{\partial^2 C(x_A,\xp,\yp,\xp',\xp')}{\partial x^i \partial x'^{i'}}\right]_{\xp = \yp}\\
+&\tr\left[(\qs + m)R_{\bar{q},\mu i}(m-\ps)\gamma^0 R'^\dag_{g,\mu' i'}\gamma^0\right]\left[\frac{\partial^2 C(x_A,\xp,\yp,\xp',\xp')}{\partial y^i \partial x'^{i'}}\right]_{\xp = \yp}\\
+&\tr\left[(\qs + m)R_{g,\mu i}(m-\ps)\gamma^0 R'^\dag_{q,\mu' i'}\gamma^0\right]\left[\frac{\partial^2 C(x_A,\xp,\xp,\yp',\xp')}{\partial x^i \partial x'^{i'}}\right]_{\xp' = \yp'}\\
+&\tr\left[(\qs + m)R_{g,\mu i}(m-\ps)\gamma^0 R'^\dag_{\bar{q},\mu' i'}\gamma^0\right]\left[\frac{\partial^2 C(x_A,\xp,\xp,\yp',\xp')}{\partial x^i \partial y'^{i'}}\right]_{\xp' = \yp'}\\
+&\tr\left[(\qs + m)R_{g,\mu i}(m-\ps)\gamma^0 R'^\dag_{g,\mu' i'}\gamma^0\right]\left[\frac{\partial^2 C(x_A,\xp,\xp,\xp',\xp')}{\partial x^i \partial x'^{i'}}\right]\Biggr\}\,.\\
\end{split}
\end{equation}

The gradients of $C(\xp,\yp,\yp'\xp')$ are parametrized in terms of
six transverse momentum dependent gluon distribution
functions as follows (see
Eqs.~(28)-(34) in \cite{Akcakaya:2012si})
\begin{equation}
\begin{split}
&\int_{\xp \xp'} e^{i\kAp\cdot(\xp - \xp')}\left[\frac{\partial^2 C(x_A,\xp,\yp,\yp',\xp')}{\partial x^i \partial x'^{i'}}\right]_{\xp = \yp, \xp' =  \yp'}\\
& = \int_{\xp \xp'} e^{i\kAp\cdot(\xp - \xp')}\left[\frac{\partial^2 C(x_A,\xp,\yp,\yp',\xp')}{\partial y^i \partial y'^{i'}}\right]_{\xp = \yp, \xp' =  \yp'}\\
&  \equiv \frac{1}{2}\delta^{ii'}F_1(x_A,k_{2\perp}^2) + 
\left(\frac{k_{2\perp}^ik_{2\perp}^{i'}}{k_{2\perp}^2} - \frac{1}{2}\delta^{ii'}\right)H_1(x_A,k_{2\perp}^2)\,,
\end{split}
\label{eq:F1H1}
\end{equation}
\begin{equation}
\begin{split}
&\int_{\xp \xp'} e^{i\kAp\cdot(\xp - \xp')}\left[\frac{\partial^2 C(x_A,\xp,\yp,\yp',\xp')}{\partial x^i \partial y'^{i'}}\right]_{\xp = \yp, \xp' =  \yp'}\\
& = \int_{\xp \xp'} e^{i\kAp\cdot(\xp - \xp')}\left[\frac{\partial^2 C(x_A,\xp,\yp,\yp',\xp')}{\partial y^i \partial x'^{i'}}\right]_{\xp = \yp, \xp' =  \yp'}\\
&  \equiv \frac{1}{2}\delta^{ii'}F_2(x_A,k_{2\perp}^2) + \left(\frac{k_{2\perp}^ik_{2\perp}^{i'}}{k_{2\perp}^2} - \frac{1}{2}\delta^{ii'}\right)H_2(x_A,k_{2\perp}^2)\,.
\end{split}
\label{eq:F2H2}
\end{equation}
\begin{equation}
\begin{split}
&\int_{\xp \xp'} e^{i\kAp\cdot(\xp - \xp')}\left[\frac{\partial^2 C(x_A,\xp,\yp,\xp',\xp')}{\partial x^i \partial x'^{i'}}\right]_{\xp = \yp} = \int_{\xp \xp'} e^{i\kAp\cdot(\xp - \xp')}\left[\frac{\partial^2 C(x_A,\xp,\yp,\xp',\xp')}{\partial y^i \partial x'^{i'}}\right]_{\xp = \yp}\\
&=\int_{\xp \xp'} e^{i\kAp\cdot(\xp - \xp')}\left[\frac{\partial^2 C(x_A,\xp,\xp,\yp',\xp')}{\partial x^i \partial x'^{i'}}\right]_{\xp' = \yp'} = \int_{\xp \xp'} e^{i\kAp\cdot(\xp - \xp')}\left[\frac{\partial^2 C(x_A,\xp,\xp,\yp',\xp')}{\partial x^i \partial y'^{i'}}\right]_{\xp' = \yp'} \\ & = \frac{1}{2}\int_{\xp \xp'} e^{i\kAp\cdot(\xp - \xp')}\left[\frac{\partial^2 C(x_A,\xp,\xp,\xp',\xp')}{\partial x^i \partial x'^{i'}}\right]\equiv \frac{1}{2}\delta^{ii'}F_3(x_A,k_{2\perp}^2) + \left(\frac{k_{2\perp}^ik_{2\perp}^{i'}}{k_{2\perp}^2} - \frac{1}{2}\delta^{ii'}\right)H_3(x_A,k_{2\perp}^2)\,.
\end{split}
\label{eq:F3H3}
\end{equation}

The cross section becomes
\begin{equation}
\begin{split}
&\frac{d\sigma}{d^2 \pp d^2 \qp d^2 \kgp d\eta_p d\eta_q d\eta_{k_\gamma}} = \frac{\alpha_e\alpha_S q_f^2}{256 \pi^8 N_c(N_c^2 - 1)}\int_{\khp \kAp}(2\pi)^2 \delta^{(2)}(\Pp - \khp - \kAp) \frac{\varphi_p(x_p,k_{1\perp}^2)}{k_{1\perp}^2}\\
&\Biggl\{\left(\tau_{qq,ii'}+\tau_{\bar{q}\bar{q},ii'}\right)\left[\frac{1}{2}\delta^{ii'}F_1(x_A,k_{2\perp}^2) + \left(\frac{k_{2\perp}^ik_{2\perp}^{i'}}{k_{2\perp}^2} - \frac{1}{2}\delta^{ii'}\right)H_1(x_A,k_{2\perp}^2)\right]\\
&+(\tau_{q\bar{q},ii'} + \tau_{\bar{q}q,ii'})\left[\frac{1}{2}\delta^{ii'}F_2(x_A,k_{2\perp}^2) + \left(\frac{k_{2\perp}^ik_{2\perp}^{i'}}{k_{2\perp}^2} - \frac{1}{2}\delta^{ii'}\right)H_2(x_A,k_{2\perp}^2)\right]\\
&+\left(\tau_{qg,ii'}+\tau_{\bar{q}g,ii'} + \tau_{gq,ii'}+\tau_{g\bar{q},ii'} + 2\tau_{gg,ii'}\right)\left[\frac{1}{2}\delta^{ii'}F_3(x_A,k_{2\perp}^2) + \left(\frac{k_{2\perp}^ik_{2\perp}^{i'}}{k_{2\perp}^2} - \frac{1}{2}\delta^{ii'}\right)H_3(x_A,k_{2\perp}^2)\right]\Biggr\}\,,\\
\end{split}
\label{eq:csi1}
\end{equation}
where
\begin{equation}
\tau_{ab,ii'} \equiv g^{\mu\mu'}\tr\left[(\qs + m)R_{a,\mu i}(m-\ps)\gamma^0 R'^\dag_{b,\mu' i'}\gamma^0\right] \,, \qquad a,b = q,\bar{q},g\,.
\end{equation}

The $F_i$ and $H_i$ distributions, introduced in
  Eqs.~\eqref{eq:F1H1}-\eqref{eq:F3H3}, have known operator
  definitions; see e.~g.~\cite{Akcakaya:2012si} where a direct
  relationships to the distributions in the conventional transverse
  momentum dependent (TMD) formalism discussed in
  \cite{Dominguez:2011wm,Bomhof:2006dp} was
  given. Ref.~\cite{Akcakaya:2012si} also established a set of linear
  relationship to the various dipole and Weizs\"acker-Williams
  gluon distributions \cite{Akcakaya:2012si}, see also
  \cite{Marquet:2017xwy}. In addition, we provide explicit expressions
  in terms of the BK two-point function of $A^+$ in
  App.~\ref{sec:F_H_distributions}.

\section{Jet-photon correlations in the collinear approximation}
\label{sec:jetgamma}

In the following we will pick up configurations where
  both the photon $\kgp$ and the antiquark (we arbitrarily choose to
  this to be the jet in this process) momentum $\pp$ are hard, while
  the left-over transverse momentum of the quark is much smaller, $\qp
  \ll \kgp$. The contributions where the photon is emitted within a
  jet can be suppressed by discarding events with a large
  transverse jet energy in a cone around the photon
  \cite{Frixione:1998jh}. The remaining configurations are in fact
  dominated by collinear photon emissions from a quark line after the
  interaction with the nuclei.
For simplicity, we will also take the collinear limit on the
proton side: $\khp \to 0$.  The cross section becomes
\begin{equation}
\begin{split}
&\frac{d\sigma}{d^2 \pp d^2 \qp d^2 \kgp d\eta_p d\eta_q d\eta_{k_\gamma}} = \frac{\alpha_e\alpha_S q_f^2}{512 \pi^6 N_c(N_c^2 - 1)}x_p f_{g,p}(x_p,P_\perp^2)\\
&\times\Biggl\{\left(\tau^{\rm coll}_{qq,ii'}+\tau^{\rm coll}_{\bar{q}\bar{q},ii'}\right)\left[\frac{1}{2}\delta^{ii'}F_1(x_A,P_\perp^2) + \left(\frac{P^i P^{i'}}{P_\perp^2} - \frac{1}{2}\delta^{ii'}\right)H_1(x_A,P_\perp^2)\right]\\
&+(\tau^{\rm coll}_{q\bar{q},ii'} + \tau^{\rm coll}_{\bar{q}q,ii'})\left[\frac{1}{2}\delta^{ii'}F_2(x_A,P_\perp^2) + \left(\frac{P^i P^{i'}}{P_\perp^2} - \frac{1}{2}\delta^{ii'}\right)H_2(x_A,P_\perp^2)\right]\\
&+\left(\tau^{\rm coll}_{qg,ii'}+\tau^{\rm coll}_{\bar{q}g,ii'} + \tau^{\rm coll}_{gq,ii'}+\tau^{\rm coll}_{g\bar{q},ii'} + 2\tau^{\rm coll}_{gg,ii'}\right)\left[\frac{1}{2}\delta^{ii'}F_3(x_A,P_\perp^2) + \left(\frac{P^i P^{i'}}{P_\perp^2} - \frac{1}{2}\delta^{ii'}\right)H_3(x_A,P_\perp^2)\right]\Biggr\}\,,\\
\end{split}
\end{equation}
where
\begin{equation}
\begin{split}
&\tau^{\rm coll}_{ab,ii'}\equiv g^{\mu\mu'}\delta^{jj'}\tr\left[(\qs + m)R_{a,\mu ij}(m-\ps)\gamma^0 R'^\dag_{b,\mu' i' j'}\gamma^0\right] \,, \qquad a,b = q,\bar{q}\,,\\
&\tau^{\rm coll}_{ga,ii'}\equiv g^{\mu\mu'}\tr\left[(\qs + m)R_{g,\mu}(m-\ps)\gamma^0 R'^\dag_{a,\mu' i' i}\gamma^0\right] \,, \qquad a = q,\bar{q}\,,\\
&\tau^{\rm coll}_{ag,ii'}\equiv g^{\mu\mu'}\tr\left[(\qs + m)R_{a,\mu i i'}(m-\ps)\gamma^0 R'^\dag_{g,\mu'}\gamma^0\right] \,, \qquad a = q,\bar{q}\,,\\
&\tau^{\rm coll}_{gg,ii'}\equiv g^{\mu\mu'}\delta_{ii'}\tr\left[(\qs + m)R_{g,\mu}(m-\ps)\gamma^0 R'^\dag_{g,\mu'}\gamma^0\right] \,.
\end{split}
\label{eq:diractr}
\end{equation}
Here we have defined $R^{\mu i}_a \equiv k_1^j R_a^{\mu ij}$, $a = q,\qbar$ and $R_g^{\mu i} \equiv k_1^i R_g^\mu$, see Eqs.~\eqref{eq:Rs} and \eqref{eq:har1}.
Also, we introduced the collinear gluon distribution of the proton,
\begin{equation}
x_p f_{g,p}(x_p,P_\perp^2) \equiv \frac{1}{4\pi^3}\int_0^{P_\perp^2} d k_{1\perp}^2 \varphi_p(x_p,k_{1\perp}^2)\,.
\end{equation}
Note that although $P_\perp$ is much smaller than the hard scale
$\tilde{P}_\perp$ it is nevertheless on the order of the saturation
scale of the nucleus and so the collinear approximation on the proton
side should be justified~\cite{Dumitru:2001jn}.

The terms in the amplitude \eqref{eq:amp3} containing the photon-quark collinear singularity are given by $R_1^{\mu i}$, $R_3^{\mu i}$ and $R_8^{\mu i}$, see Eq.~\eqref{eq:har1}. To correctly take into account this singularity the Dirac traces \eqref{eq:diractr} must contain the diagonal parts corresponding to the matrices $R_1^{\mu i}$, $R_3^{\mu i}$ and $R_8^{\mu i}$, as well as interferences with the remaining part of the amplitude, leading to
\begin{equation}
\begin{split}
&\tau^{\rm coll}_{qq,ii'} = \tau^{\rm coll}_{33,ii'} + \tau^{\rm coll}_{34,ii'} + \tau^{\rm coll}_{43,ii'} + \tau^{\rm coll}_{35,ii'} + \tau^{\rm coll}_{53,ii'}\,,\\
&\tau^{\rm coll}_{\qbar \qbar,ii'} = \tau^{\rm coll}_{88,ii'} + \tau^{\rm coll}_{87,ii'} + \tau^{\rm coll}_{78,ii'} + \tau^{\rm coll}_{86,ii'} + \tau^{\rm coll}_{68,ii'}\,,\\
&\tau^{\rm coll}_{q \qbar,ii'} = \tau^{\rm coll}_{36,ii'} + \tau^{\rm coll}_{37,ii'} + \tau^{\rm coll}_{38,ii'} + \tau^{\rm coll}_{48,ii'} + \tau^{\rm coll}_{58,ii'}\,,\\
&\tau^{\rm coll}_{\qbar q,ii'} = \tau^{\rm coll}_{63,ii'} + \tau^{\rm coll}_{73,ii'} + \tau^{\rm coll}_{83,ii'} + \tau^{\rm coll}_{84,ii'} + \tau^{\rm coll}_{85,ii'}\,,\\
&\tau^{\rm coll}_{gq,ii'} = \tau^{\rm coll}_{13,ii'} + \tau^{\rm coll}_{14,ii'} + \tau^{\rm coll}_{15,ii'} + \tau^{\rm coll}_{23,ii'}\,,\\
&\tau^{\rm coll}_{g\qbar,ii'} = \tau^{\rm coll}_{16,ii'} + \tau^{\rm coll}_{17,ii'} + \tau^{\rm coll}_{18,ii'} + \tau^{\rm coll}_{28,ii'}\,,\\
&\tau^{\rm coll}_{qg,ii'} = \tau^{\rm coll}_{31,ii'} + \tau^{\rm coll}_{41,ii'} + \tau^{\rm coll}_{51,ii'} + \tau^{\rm coll}_{32,ii'}\,,\\
&\tau^{\rm coll}_{\qbar g,ii'} = \tau^{\rm coll}_{61,ii'} + \tau^{\rm coll}_{71,ii'} + \tau^{\rm coll}_{81,ii'} + \tau^{\rm coll}_{82,ii'}\,,\\
&\tau^{\rm coll}_{gg,ii'} = \tau^{\rm coll}_{11,ii'} + \tau^{\rm coll}_{12,ii'} + \tau^{\rm coll}_{21,ii'}\,.\\
\end{split}
\label{eq:taucoll}
\end{equation}
Here we have defined
\begin{equation}
\begin{split}
&\tau^{\rm coll}_{\alpha\beta,ii'}\equiv g^{\mu\mu'}\delta^{jj'}\tr\left[(\qs + m)R_{\alpha,\mu ij}(m-\ps)\gamma^0 R'^\dag_{\beta,\mu' i' j'}\gamma^0\right] \,, \qquad \alpha,\beta = 3,\dots,8\,,\\
&\tau^{\rm coll}_{\alpha\beta,ii'}\equiv g^{\mu\mu'}\tr\left[(\qs + m)R_{\alpha,\mu}(m-\ps)\gamma^0 R'^\dag_{\beta,\mu' i' i}\gamma^0\right] \,, \qquad \alpha = 1,2 \,, \quad  \beta = 3,\dots,8\,,\\
&\tau^{\rm coll}_{\alpha\beta,ii'}\equiv g^{\mu\mu'}\tr\left[(\qs + m)R_{\alpha,\mu i i'}(m-\ps)\gamma^0 R'^\dag_{\beta,\mu'}\gamma^0\right] \,, \qquad \alpha = 3 \,, \dots,8 \,, \quad \beta = 1,2 \,,\\
&\tau^{\rm coll}_{\alpha\beta,ii'}\equiv g^{\mu\mu'}\delta_{ii'}\tr\left[(\qs + m)R_{\alpha,\mu}(m-\ps)\gamma^0 R'^\dag_{\beta,\mu'}\gamma^0\right] \,, \qquad \alpha,\beta = 1,2 \,,
\end{split}
\end{equation}
where $R^{\mu i}_\alpha \equiv k_1^j R_\alpha^{\mu ij}$, $\alpha = 3,\dots,8$, and $R_\alpha^{\mu i} \equiv k_1^i R_\alpha^\mu$, $\alpha = 1,2$, see Eqs.~\eqref{eq:Rs}.

\subsection{Integration over the photon-quark collinear singularity}

We introduce the photon momentum fraction
\begin{equation}
z \equiv \frac{k_\gamma^+}{k_\gamma^+ + q^+}\,,
\end{equation}
and separate out the collinear singularity in the Dirac traces
\begin{equation}
\tau^{\rm coll}_{ab, ii'}(\wperp) \equiv \frac{\tau^{\rm coll, \, reg}_{ab, ii'}(\wperp)}{[z\wperp - \kgp]^2}\,.
\end{equation}
The integral over the transverse phase space $\qp$ is dominated by the
singularity $\wperp \equiv \qp + \kgp = \kgp/z$. The integral is
calculated by expanding the integrand around this singularity and then
integrating over it.

A generic integral is
\begin{equation}
I = \int_{\qp} \frac{F(\wperp,\pp)}{(z\wperp - \kgp)^2 (\wperp + \pp)^2} = \int_{\wperp} \frac{F(\wperp,\pp)}{(z\wperp - \kgp)^2 (\wperp + \pp)^2}\,,
\label{eq:geni}
\end{equation}
where in the second line we shifted the integration $\qp \to \wperp$. $F(\wperp,\pp)$ is a function containing the hard parts that potentially depend on $\wperp$ and $\pp$ as well as the distribution function that depends on $\Pp = \wperp + \pp$. In \eqref{eq:geni} we have separated out explicitly the perturbative tail $\sim 1/P_\perp^2$ of the distribution functions. 
Expanding $F(\wperp, \pp)$ around the collinear singularity we obtain
\begin{equation}
I  \simeq F(\kgp/z,\pp) \int_{\wperp} \frac{1}{(z\wperp - \kgp)^2 (\wperp + \pp)^2} = \frac{1}{2\pi}\frac{F(\kgp/z,\pp)}{(\kgp + z\pp)^2}\left\{-\frac{1}{\epsilon}+\gamma_E + \log\left[\frac{(\kgp + z\pp)^2}{4\pi z^2 \mu^2}\right] + O(\epsilon)\right\}\,,
\end{equation}
where in the second line we evaluated the integral in $d$-dimensions and expanded around $d = 2- 2\epsilon$. Here $\mu^2$ is the renormalization scale. Subtracting the $1/\epsilon$ divergence and defining $\mu^2 = \mu^2_{\msbar} e^{\gamma_E}/4\pi$, the final formula for the integral in the $\msbar$ scheme is
\begin{equation}
I_{\msbar} = \frac{1}{2\pi}\frac{F(\kgp/z,\pp)}{(\kgp + z\pp)^2}\log\left[\frac{(\kgp + z\pp)^2}{z^2 \mu_{\msbar}^2}\right]\,.
\label{eq:geni2}
\end{equation}

Using \eqref{eq:geni2} we calculate the $\qp$ integrals in the cross section as
\begin{equation}
\begin{split}
&\int_{\qp}\frac{\tau^{\rm coll}_{ab, ii'}(\wperp)}{(\wperp + \pp)^2}P_\perp^2\left[\frac{1}{2}\delta^{ii'}F_j(x_A,P_\perp^2) + \left(\frac{P^i P^{i'}}{P_\perp^2} - \frac{1}{2}\delta^{ii'}\right)H_j(x_A,P_\perp^2)\right] = \frac{1}{2\pi}\frac{1}{z^2}\frac{1}{(\pp + \kgp/z)^2}\log\left[\frac{(\pp + \kgp/z)^2}{\mu_{\msbar}^2}\right]\\
&\times \tau^{\rm coll, \, reg}_{ab, ii'}(\kgp/z)(\pp + \kgp/z)^2 \Bigg[\frac{1}{2}\delta^{ii'}F_j(x_A,(\pp + \kgp/z)^2)\\
& + \left(\frac{( p^i + k_\gamma^i/z) (p^{i'} + k_\gamma^{i'}/z)}{(\pp + \kgp/z)^2} - \frac{1}{2}\delta^{ii'}\right)H_j(x_A,(\pp + \kgp/z)^2)\Bigg]\,.
\end{split}
\label{eq:geni3}
\end{equation}

\subsection{Cross section in the correlation limit}
\label{sec:CorrelLimit}

We calculate $\tau^{\rm coll, \, reg}_{ab, ii'}(\kgp/z)$ assuming that
the total momentum of the final state in the correlation limit, $\pp +
\kgp/z$, is small. This follows from the fact that the hard factors
have been expanded around $\khp, \kAp \to 0$ and from momentum
conservation. We introduce new variables
\begin{equation}
\Qp \equiv \kgp + \pp \,, \qquad \Ppt \equiv \frac{1}{2}\left(\pp - \kgp\right) \,.
\end{equation}
Below we will consider the limit where $Q_\perp \ll \tilde{P}_\perp$. Since this means that both $Q_\perp$ and $|\pp + \kgp/z|$ are small, we are considering events where the photon picks up most of the energy of its parent quark, so that $z$ is close to unity.

The explicit expressions for the hard factors are
\begin{equation}
\begin{split}
&\tau^{\rm coll, \, reg}_{qq,ii'}(\kgp/z) + \tau^{\rm coll, \, reg}_{\qbar\qbar,ii'}(\kgp/z) = \frac{2 \zeta z^2 (1-z) (1+z)^4}{(\zeta+z)^4} \frac{1}{\tilde{P}_\perp^4}\left((\zeta^2 + z^2)\delta_{ii'} - \frac{4\zeta z(\zeta-z)^2}{(\zeta+z)^2} \frac{\tilde{P}_i \tilde{P}_{i'}}{\tilde{P}_\perp^2}\right) \frac{1+(1-z)^2}{z}\,,\\
&\tau^{\rm coll, \, reg}_{q\qbar,ii'}(\kgp/z) + \tau^{\rm coll, \, reg}_{\qbar q,ii'}(\kgp/z) =  -\frac{8 \zeta^2 z^3 (1-z) (1+z)^4(\zeta-z)^2}{(\zeta+z)^6} \frac{\tilde{P}_i \tilde{P}_{i'}}{\tilde{P}_\perp^6} \frac{1+(1-z)^2}{z}\,,\\
&\tau^{\rm coll, \, reg}_{g q,ii'}(\kgp/z) + \tau^{\rm coll, \, reg}_{qg,ii'}(\kgp/z) + \tau^{\rm coll, \, reg}_{g \qbar,ii'}(\kgp/z) + \tau^{\rm coll, \, reg}_{\qbar g,ii'}(\kgp/z) + 2\tau^{\rm coll, \, reg}_{gg,ii'}(\kgp/z)\\
& =  -\frac{4 \zeta^2 z^3 (1-z) (1+z)^4}{(\zeta+z)^6} \frac{1}{\tilde{P}_\perp^4}\left((\zeta^2 + z^2)\delta_{ii'} - 2(\zeta-z)^2 \frac{\tilde{P}_i \tilde{P}_{i'}}{\tilde{P}_\perp^2}\right) \frac{1+(1-z)^2}{z}\,,\\
\end{split}
\label{eq:hard}
\end{equation}
where we see explicitly the photon splitting function $(1+(1-z)^2)/z$.
We also introduced the abbreviation
\begin{equation}
\zeta \equiv \frac{k_\gamma^+}{p^+} = \frac{k_{\gamma\perp}}{p_\perp}e^{\eta_{k_\gamma}-\eta_p}\,.
\end{equation}
In terms of $\Qp$ and $\Ppt$ the transverse momentum in the nuclear
gluon distributions is
\begin{equation}
\pp + \frac{1}{z}\kgp = -\frac{1-z}{z}\Ppt + \frac{1+z}{2z}\Qp\,.
\end{equation}

Next, we multiply the transverse projector in Eq.~\eqref{eq:geni3} with $\tilde{P}_i \tilde{P}_{i'}$ from \eqref{eq:hard}
\begin{equation}
\begin{split}
\left(\frac{( p^i + k_\gamma^i/z) (p^{i'} + k_\gamma^{i'}/z)}{(\pp + \kgp/z)^2} - \frac{1}{2}\delta^{ii'}\right)\frac{\tilde{P}_i \tilde{P}_{i'}}{\tilde{P}_\perp^2} & = \frac{\left(-\frac{1-z}{z}\tilde{P}_\perp^2 + \frac{1+z}{2z}\Qp\cdot\Ppt\right)^2}{\left(-\frac{1-z}{z}\Ppt + \frac{1+z}{2z}\Qp\right)^2 \tilde{P}_\perp^2}-\frac{1}{2}\\
& \simeq \frac{1}{2} - \frac{(1+z)^2}{4(1-z)^2}\frac{Q_\perp^2}{\tilde{P}_\perp^2} + \frac{(1+z)^2}{4(1-z)^2}\frac{(\Qp\cdot\Ppt)^2}{\tilde{P}_\perp^4}\,.
\end{split}
\end{equation}
We took the correlation limit $Q_\perp \ll \tilde{P}_\perp$ in the
last line but in the cross section below we will keep the full
expression.  The full cross section, integrated over the quark
momentum $\qp$ is
\begin{equation}
\begin{split}
&\frac{d\sigma}{d^2 \Ppt d^2 \Qp d\eta_p d\eta_{k_\gamma} dz} =
  \frac{\alpha_e\alpha_S q_f^2}{64\pi^4 N_c(N_c^2 -1)} x_p
  f_{g,p}\left(x_p,\left(-\frac{1-z}{z}\Ppt +
  \frac{1+z}{2z}\Qp\right)^2\right)\\
&\times
  \frac{1}{2\pi}\frac{1+(1-z)^2}{z} \log\left[\frac{\left(-\frac{1-z}{z}\Ppt
      + \frac{1+z}{2z}\Qp\right)^2}{\Lambda_{\msbar}^2}\right]\\
&\times\frac{\zeta(1+z)^4}{z(\zeta+z)^6}\frac{1}{\tilde{P}_\perp^4}\Bigg\{(\zeta^4 + 6\zeta^2 z^2 + z^4)F_1\left(x_A,\left(-\frac{1-z}{z}\Ppt + \frac{1+z}{2z}\Qp\right)^2\right)\\
& - 2\zeta z(\zeta-z)^2 F_2\left(x_A,\left(-\frac{1-z}{z}\Ppt + \frac{1+z}{2z}\Qp\right)^2\right) - 4\zeta^2 z^2 F_3\left(x_A,\left(-\frac{1-z}{z}\Ppt + \frac{1+z}{2z}\Qp\right)^2\right)\\
& + 4\zeta z(\zeta-z)^2\left[\frac{\frac{(1-z)^2}{z^2}\tilde{P}_\perp^2 + \frac{(1+z)^2}{4z^2}\frac{(\Qp\cdot \Ppt)^2}{\tilde{P}_\perp^2} - \frac{1-z^2}{z^2}\Qp\cdot \Ppt}{\left(-\frac{1-z}{z}\Ppt + \frac{1+z}{2z}\Qp\right)^2}-\frac{1}{2}\right]\\
&\times\Bigg[-H_1\left(x_A,\left(-\frac{1-z}{z}\Ppt + \frac{1+z}{2z}\Qp\right)^2\right)-H_2\left(x_A,\left(-\frac{1-z}{z}\Ppt + \frac{1+z}{2z}\Qp\right)^2\right)\\
&+H_3\left(x_A,\left(-\frac{1-z}{z}\Ppt + \frac{1+z}{2z}\Qp\right)^2\right)\Bigg]\Bigg\}\,.
\end{split}
\label{eq:Xsection_allQ}
\end{equation}
Here we have added the contribution to the cross section coming from the fragmentation photons proportional to the quark-photon fragmentation function
\begin{equation}
D_{q\to\gamma}(z,\mu^2) =
\frac{1}{2\pi}\frac{1+(1-z)^2}{z} \log
\left(\frac{\mu^2}{\Lambda_{\msbar}^2}\right) ~.
\end{equation}
That way, the cross section is independent of the arbitrary
renormalization scale $\mu^2$.

It follows from \eqref{eq:His} that the combination of linearly
polarized gluon distributions that appears in the cross section vanishes:
\begin{equation}
-H_1(x_A,k_\perp^2) -H_2(x_A,k_\perp^2) + H_3(x_A,k_\perp^2)  = 0\,,
\end{equation}
We now expand the cross section in powers of $Q_\perp$\footnote{In
  fact, the expansion is in powers of $\frac{1+z}{1-z}Q_\perp/
  \tilde{P}_\perp$. Hence, for large $z$ the expansions
  given below apply only for rather small values of
  $Q_\perp/\tilde{P}_\perp$. For larger $Q_\perp/\tilde{P}_\perp$ one
  could compute the angular correlations numerically from Eq.~(\ref{eq:Xsection_allQ}).}:
\begin{equation}
\begin{split}
&\frac{d\sigma}{d^2 \Ppt d^2 \Qp d\eta_p d\eta_{k_\gamma} dz} = \frac{\alpha_e\alpha_S q_f^2}{64\pi^4 N_c(N_c^2 -1)} \frac{1}{z(1-z)} x_p f_{g,p}\left(x_p,\frac{(1-z)^2}{z^2}\tilde{P}_\perp^2\right)\frac{1}{2\pi}\frac{1+(1-z)^2}{z}\log\left[\frac{(1-z)^2\tilde{P}_\perp^2}{z^2\Lambda_{\msbar}^2}\right]\\
&\times\frac{\zeta(1-z)(1+z)^4}{(\zeta+z)^6}\frac{1}{\tilde{P}_\perp^4}\sum_{n=0} ^\infty \frac{1}{n!}\Bigg\{(\zeta^4 + 6\zeta^2 z^2 + z^4)\left[\frac{\partial^n F_1}{\partial Q^{i_1}\dots \partial Q^{i_n}}\right]_{\Qp = 0}\\
& - 2\zeta z(\zeta-z)^2 \left[\frac{\partial^n F_2}{\partial Q^{i_1}\dots \partial Q^{i_n}}\right]_{\Qp = 0} - 4\zeta^2 z^2 \left[\frac{\partial^n F_3}{\partial Q^{i_1}\dots \partial Q^{i_n}}\right]_{\Qp = 0}\Bigg\}Q^{i_1}\dots Q^{i_n}\,.
\end{split}
\end{equation}
In this expansion we have neglected the dependence of $x_p$ and $x_A$
on $Q_\perp$ in the proton and the nuclear distributions. In both
cases the derivatives will bring energy denominators that suppress
such contributions in the high energy limit. In addition, the
derivative of the proton distribution over $\log 1/x_p$ would be
$\alpha_S$ suppressed by virtue of the Balitsky-Fadin-Kuraev-Lipatov (BFKL) equation \cite{BFKL}. The nuclear
distributions evolve according to the BK equation which leads
to an even stronger suppression for transverse momenta of order (or
less than) the saturation scale of the nucleus.

Evaluating the derivatives of any of the $F_j$ we get
\begin{equation}
\left[\frac{\partial F_j}{\partial Q^{i_1}\dots \partial Q^{i_n}}\right]_{\Qp = 0} Q^{i_1}\dots Q^{i_n} = (-i)^n\frac{(1+z)^n}{(2z)^n} (\pi R_A^2) \int_{\xp} (\Qp\cdot \xp)^n \exp\left(i\frac{1-z}{z}\Ppt\cdot \xp\right) \tilde{F}_j(x_A,x_\perp^2)\,,
\label{eq:fjder}
\end{equation}
where $\tilde{F}_j(x_A,x_\perp^2)$ is the Fourier transform of the distribution functions $F_j(x_A,k_\perp^2)$.
For $n=1,2$ we find
\begin{equation}
\left[\frac{\partial F_j}{\partial Q^{i_1}}\right]_{\Qp = 0} Q^{i_1} = \cos\phi \frac{1+z}{2z} \frac{z}{1-z} \frac{Q_\perp}{\tilde{P}_\perp}F^{(1,1)}_j\left(x_A,\frac{(1-z)^2}{z^2}\tilde{P}_\perp^2\right) \,,
\end{equation}
and
\begin{equation}
\left[\frac{\partial^2 F_j}{\partial Q^{i_1}\partial
    Q^{i_2}}\right]_{\Qp = 0} Q^{i_1}Q^{i_2} = \frac{1}{2}\frac{(1+z)^2}{4z^2} \frac{z^2}{(1-z)^2}\frac{Q_\perp^2}{\tilde{P}_\perp^2}
\left[-F^{(0,2)}_j\left(x_A,\frac{(1-z)^2}{z^2}\tilde{P}_\perp^2\right) + \cos2\phi \,
F^{(2,2)}_{j}\left(x_A,\frac{(1-z)^2}{z^2}\tilde{P}_\perp^2\right)\right]\,,
\end{equation}
where
\begin{equation}
F^{(a,b)}_j\left(x_A,k_\perp^2\right) \equiv (\pi R_A^2) \, 2\pi\int_0^\infty x_\perp dx_\perp  J_a\left(k_\perp x_\perp\right) (x_\perp k_\perp)^b \tilde{F}_j(x_\perp^2) \,,
\end{equation}
and where $\cos\phi \equiv \Qp\cdot \Ppt/(Q_\perp\tilde{P}_\perp)$. Note that $F^{(0,0)}_j\left(x_A,k_\perp^2\right) = F_j\left(x_A,k_\perp^2\right)$.
The cross section, expanded up to the order $Q_\perp^2/P_\perp^2$ is
\begin{equation}
\begin{split}
&\frac{d\sigma}{d^2 \Ppt d^2 \Qp d\eta_p d\eta_{k_\gamma} dz} =
  \frac{\alpha_e\alpha_S q_f^2}{64\pi^4 N_c(N_c^2 -1)} x_p
  f_{g,p}\left(x_p,\frac{(1-z)^2}{z^2} \tilde{P}_\perp^2\right)
  \frac{1}{2\pi}
  \frac{1+(1-z)^2}{z}\log\left[\frac{(1-z)^2\tilde{P}_\perp^2}{z^2\Lambda_{\msbar}^2}\right]\\
  &\times\frac{\zeta(1+z)^4}{z(\zeta+z)^6}\frac{1}{\tilde{P}_\perp^4}\Bigg\{(\zeta^4
  + 6\zeta^2 z^2 +
  z^4)\left[F_1\left(x_A,\frac{(1-z)^2}{z^2}P_\perp^2\right)-\frac{1}{4}
    \frac{(1+z)^2}{4z^2}\frac{z^2}{(1-z)^2}\frac{Q_\perp^2}{\tilde{P}_\perp^2}F_1^{(0,2)}\left(x_A,\frac{(1-z)^2}{z^2}P_\perp^2\right)\right]\\ &
  - 2\zeta z(\zeta-z)^2
  \left[F_2\left(x_A,\frac{(1-z)^2}{z^2}P_\perp^2\right)-\frac{1}{4}
    \frac{(1+z)^2}{4z^2}\frac{z^2}{(1-z)^2}\frac{Q_\perp^2}{\tilde{P}_\perp^2}F_2^{(0,2)}\left(x_A,\frac{(1-z)^2}{z^2}P_\perp^2\right)\right]\\ &
  - 4\zeta^2 z^2
  \left[F_3\left(x_A,\frac{(1-z)^2}{z^2}P_\perp^2\right)-\frac{1}{4}
    \frac{(1+z)^2}{4z^2}\frac{z^2}{(1-z)^2}\frac{Q_\perp^2}{\tilde{P}_\perp^2}F_3^{(0,2)}\left(x_A,\frac{(1-z)^2}{z^2}P_\perp^2\right)\right]\\ &+\cos\phi\frac{1+z}{2z}\frac{z}{1-z}\frac{Q_\perp}{\tilde{P}_\perp}\Bigg[(\zeta^4
    + 6\zeta^2 z^2 +
    z^4)F_1^{(1,1)}\left(x_A,\frac{(1-z)^2}{z^2}P_\perp^2\right)-
    2\zeta
    z(\zeta-z)^2F_2^{(1,1)}\left(x_A,\frac{(1-z)^2}{z^2}P_\perp^2\right)\\ &-
    4\zeta^2 z^2
    F_3^{(1,1)}\left(x_A,\frac{(1-z)^2}{z^2}P_\perp^2\right)\Bigg]\\ &+\frac{1}{4}\cos
  2\phi\frac{(1+z)^2}{4
    z^2}\frac{z^2}{(1-z)^2}\frac{Q_\perp^2}{\tilde{P}_\perp^2}\Bigg[(\zeta^4
    + 6\zeta^2 z^2 +
    z^4)F_1^{(2,2)}\left(x_A,\frac{(1-z)^2}{z^2}P_\perp^2\right)-
    2\zeta z(\zeta-z)^2
    F_2^{(2,2)}\left(x_A,\frac{(1-z)^2}{z^2}P_\perp^2\right)\\ &-
    4\zeta^2 z^2
    F_3^{(2,2)}\left(x_A,\frac{(1-z)^2}{z^2}P_\perp^2\right)\Bigg]+O\left(Q_\perp^3/\tilde{P}_\perp^3\right)\Bigg\}\,.
\end{split}
\label{eq:xcorr}
\end{equation}

In the collinear approximation $x_{p,A}$ become
\begin{equation}
\begin{split}
& x_p\sqrt{s} = p_\perp e^{\eta_p} + zk_{\gamma\perp} e^{\eta_{k_\gamma}}\,,\\
& x_A\sqrt{s} = p_\perp e^{-\eta_p} + \frac{1}{z} k_{\gamma\perp} e^{-\eta_{k_\gamma}}\,.\\
\end{split}
\end{equation}

\subsection{Angular correlations}

We can define angular correlations via
\begin{equation}
a_n \equiv \langle \cos(n\phi)\rangle \equiv
\frac{\int_0^{2\pi}d\phi\frac{d\sigma}{d^2 \Ppt d^2 \Qp d\eta_p
    d\eta_{k_\gamma}
    dz}\cos(n\phi)}{\int_0^{2\pi}d\phi\frac{d\sigma}{d^2 \Ppt d^2 \Qp
    d\eta_p d\eta_{k_\gamma} dz}}\, ,
\end{equation}
where $\phi$ is the angle between the photon and jet transverse
momentum imbalance $\Qp$ and their average transverse momentum $\Ppt$.
In principle, all $a_n$ can be non-zero but are increasingly
suppressed like $(Q_\perp/\tilde{P}_\perp)^n$. From \eqref{eq:xcorr} we
can obtain $a_1$ and $a_2$ as follows:
\begin{equation}
a_1 = \frac{1+z}{4(1-z)}\frac{Q_\perp}{\tilde{P}_\perp}\frac{(\zeta^4 + 6\zeta^2 z^2 + z^4)F_1^{(1,1)}-2\zeta z(\zeta - z)^2 F_2^{(1,1)}-4\zeta^2 z^2 F_3^{(1,1)}}{(\zeta^4 + 6\zeta^2 z^2 + z^4)F_1-2\zeta z(\zeta - z)^2 F_2-4\zeta^2 z^2 F_3} + O\left(Q_\perp^3/\tilde{P}_\perp^3\right)\,,
\label{eq:a1}
\end{equation}
\begin{equation}
a_2 =
\frac{(1+z)^2}{32(1-z)^2}\frac{Q_\perp^2}{\tilde{P}_\perp^2}\frac{(\zeta^4
  + 6\zeta^2 z^2 + z^4)F_1^{(2,2)}-2\zeta z(\zeta - z)^2
  F_2^{(2,2)}-4\zeta^2 z^2 F_3^{(2,2)}}{(\zeta^4 + 6\zeta^2 z^2 +
  z^4)F_1-2\zeta z(\zeta - z)^2 F_2-4\zeta^2 z^2 F_3} +
O\left(Q_\perp^3/\tilde{P}_\perp^3\right)\,.
\label{eq:a2}
\end{equation}
This is the main result of this paper.
Recall that $\zeta \equiv \frac{k_\gamma^+}{p^+} =
\frac{k_{\gamma\perp}}{p_\perp}e^{\eta_{k_\gamma}-\eta_p}$ is the
ratio of photon and quark energies in the final state; while
$z$ is the fractional energy picked up by the photon from its parent
quark. Our result applies when $1-z\ll1$.
Note that the gluon distributions in the expressions above are
evaluated at the scale $(1-z)\tilde{P}_\perp/z$. Hence, these angular
moments provide insight into the transverse momentum dependence of the
gluon distributions of the target which is due to the fact that they
involve {\em another scale}, i.e.\ the non-linear (``saturation'') scale $Q_S$.

\begin{figure}
  \begin{center}
  \includegraphics[scale = 0.5]{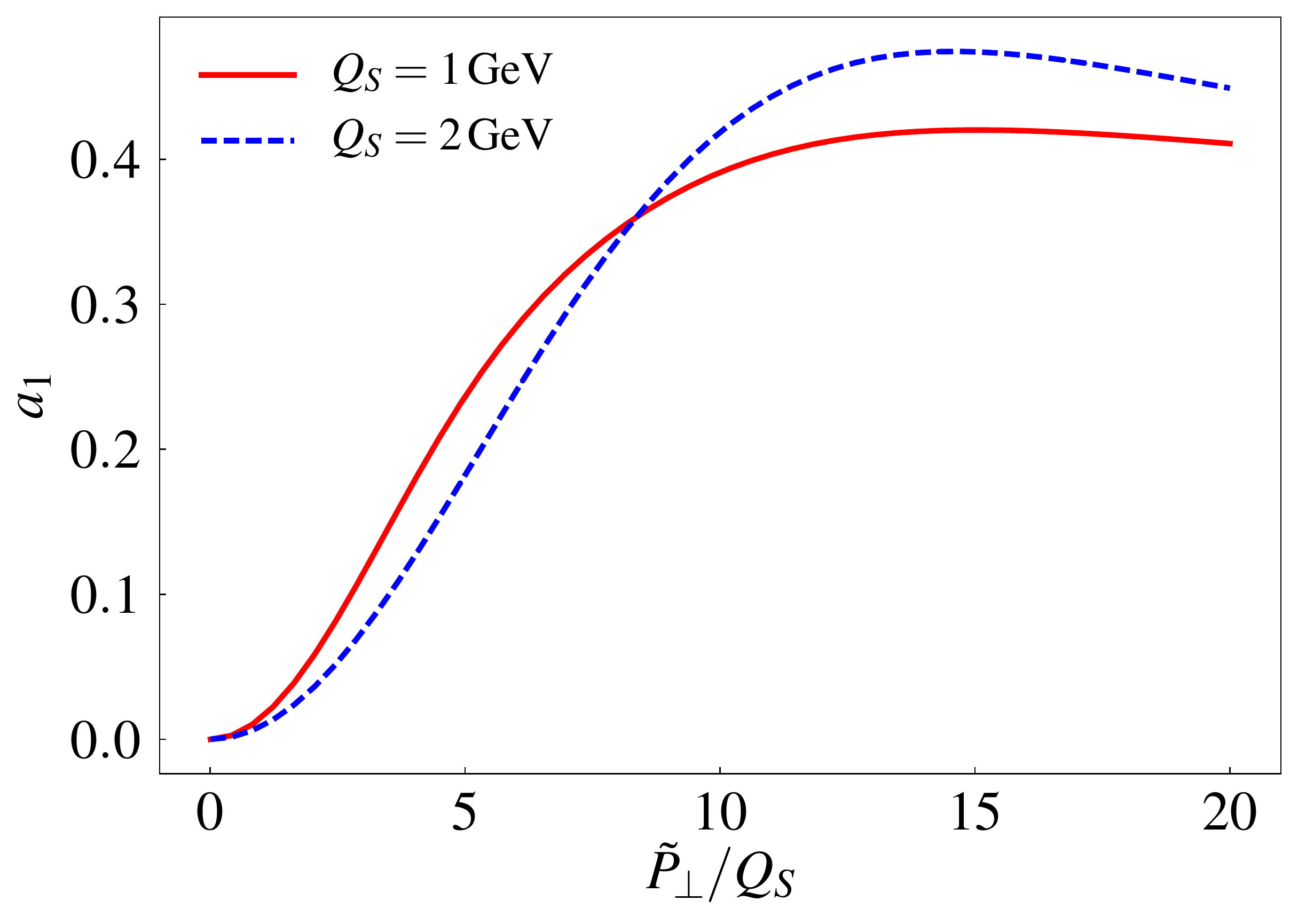}
  \end{center}
  \caption{$a_1$ for $g\to q\gamma$ to order
    $Q_\perp^2/\tilde{P}_\perp^2$ for the MV model with two different
    saturation scales. We take $Q_\perp/\tilde{P}_\perp = 0.1$, $z =
    3/4$ and $\zeta = 1$.}
  \label{fig:v1mv}
\end{figure}

\begin{figure}
  \begin{center}
  \includegraphics[scale = 0.5]{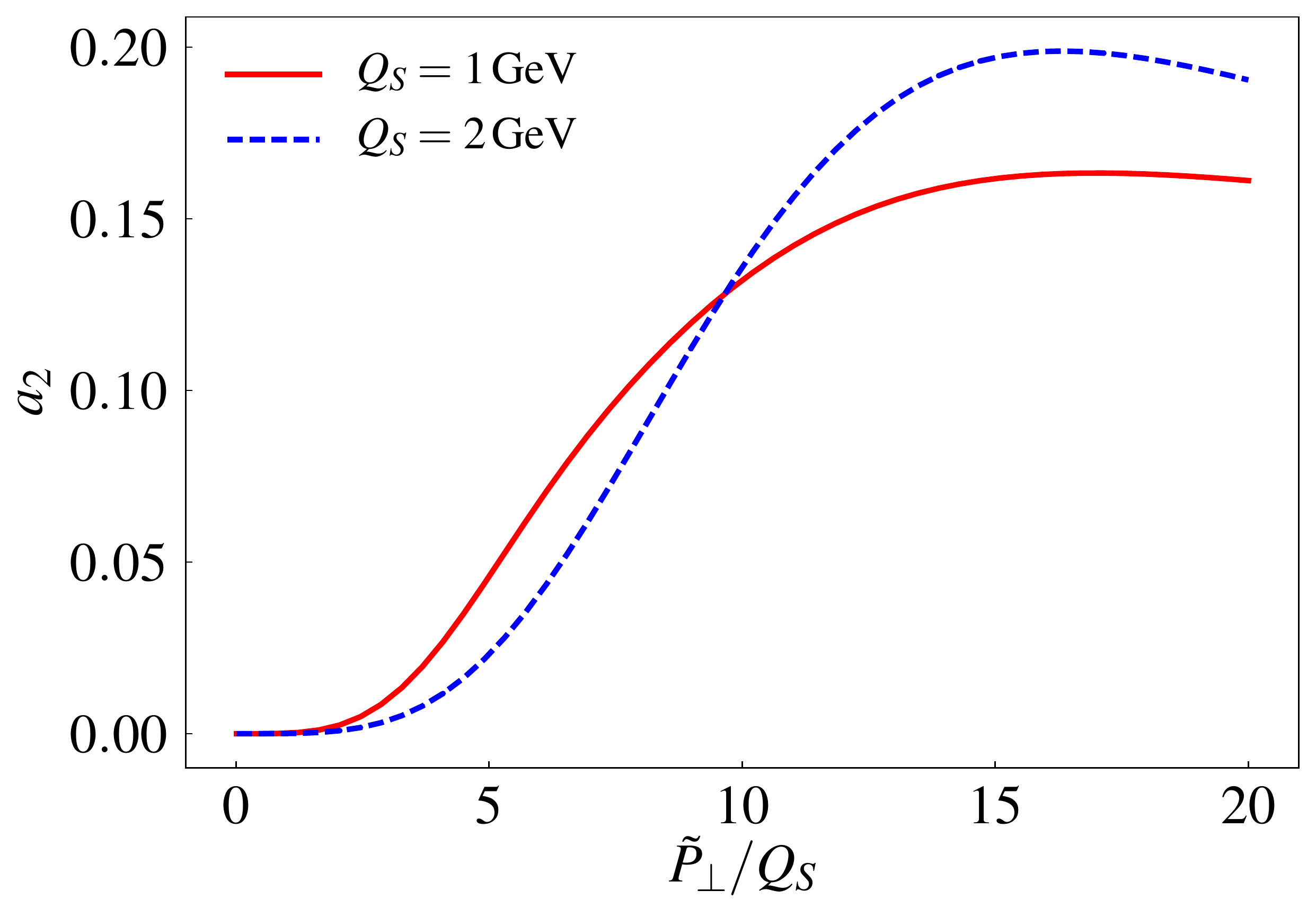}
  \end{center}
  \caption{$a_2$ for $g\to q\gamma$ to order $Q_\perp^2/\tilde{P}_\perp^2$ for the MV
    model with two different saturation scales. We take
    $Q_\perp/\tilde{P}_\perp = 0.1$, $z = 3/4$ and $\zeta = 1$.}
  \label{fig:v2mv}
\end{figure}

For the numerical results shown in Figs.~\ref{fig:v1mv} and
\ref{fig:v2mv} we have used $Q_\perp/\tilde{P}_\perp = 0.1$, $z =
3/4$, $\zeta=1$. Also, for simplicity we have computed the $F_i$ and
$F_i^{(a,b)}$ gluon distributions in the Mclerran-Venugopalan (MV) model \cite{MV} with detailed calculations in App.~\ref{sec:F_H_distributions}, while leaving a numerical
study of evolution effects for the future. Both $a_1(\tilde{P}_\perp)$
and $a_2(\tilde{P}_\perp)$ increase rapidly with transverse momentum
and reach their maximal values approximately when
$(1-z)\tilde{P}_\perp/z \approx5 Q_S$.  This maximum may disappear if
evolution effects are taken into account, similar to their effect on
the dipole scattering amplitude~\cite{BK_Cronin}. Also, we find that
$a_2$ is substantially smaller than $a_1$ due to the suppression by
one additional power of $Q_\perp/\tilde{P}_\perp$.

\begin{figure}
  \begin{center}
  \includegraphics[scale = 0.5]{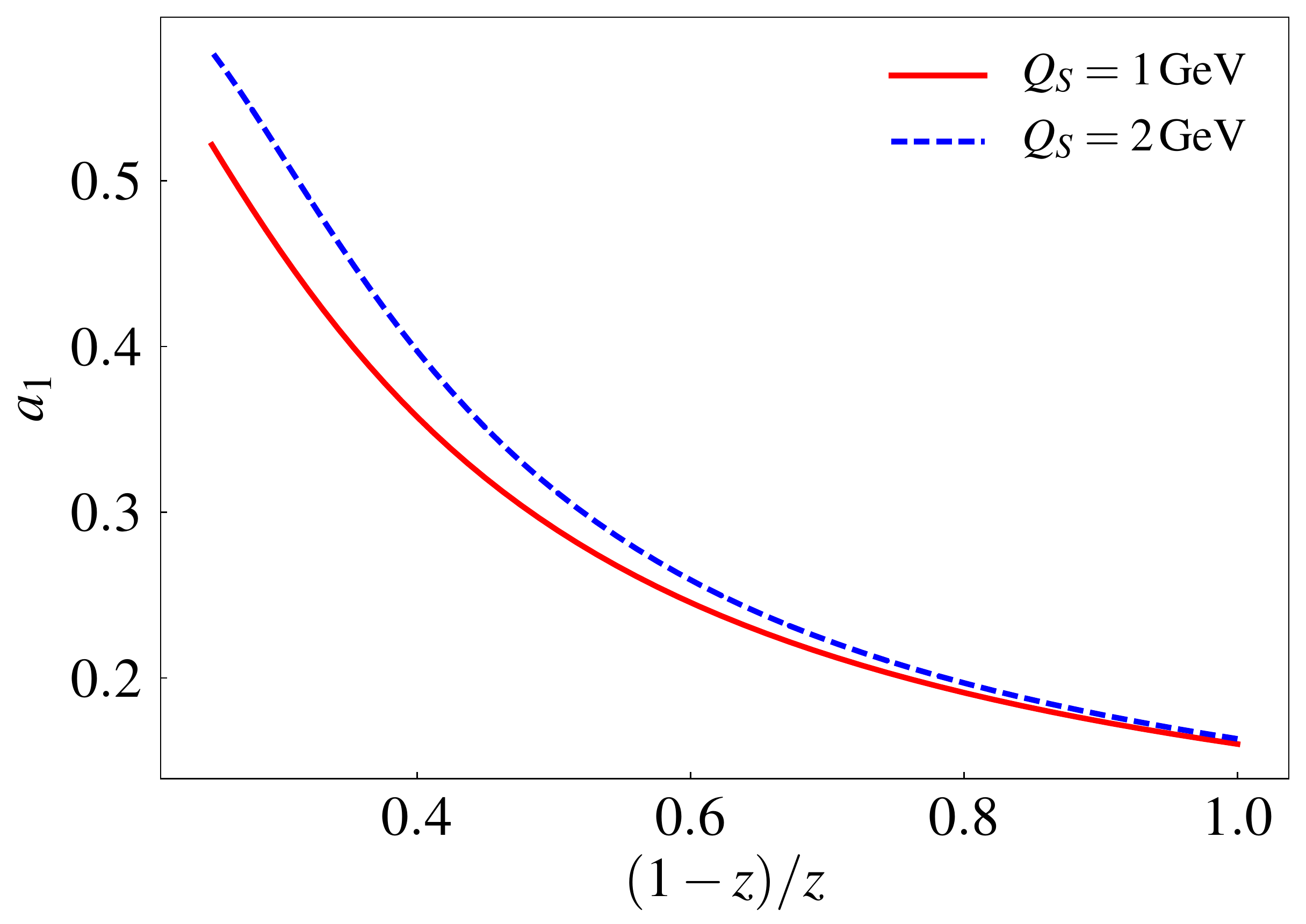}
  \end{center}
  \caption{$a_1$ for $g\to q\gamma$ to order
    $Q_\perp^2/\tilde{P}_\perp^2$ as a function of the photon
    isolation cut for the MV model with two different saturation
    scales. We take $Q_\perp/\tilde{P}_\perp = 0.1$, $\tilde{P}_\perp
    = 15Q_S$ and $\zeta =1$.}
  \label{fig:v1zmv}
\end{figure}

In Fig.~\ref{fig:v1zmv}
we plot $a_1$
as a function of the photon isolation cut $(1-z)/z$, which is the energy
fraction carried by the final state quark collinear to the photon. The
behavior of $a_2$ is very similar.  As before we used $\zeta = 1$,
$Q_\perp/\tilde{P}_\perp = 0.1$, $\tilde{P}_\perp = 15Q_S$, and we
show curves for two different values of the saturation scale. It is
interesting to observe how reducing $(1-z)/z$ increases the angular
correlations. 

We may also compute the ``angular harmonics'' given by
\begin{equation}
v_n^2 \equiv \langle \cos(n\Phi)\rangle \equiv
\frac{\int_0^{2\pi}d\phi\frac{d\sigma}{d^2 \Ppt d^2 \Qp d\eta_p
    d\eta_{k_\gamma}
    dz}\cos(n\Phi)}{\int_0^{2\pi}d\phi\frac{d\sigma}{d^2 \Ppt d^2 \Qp
    d\eta_p d\eta_{k_\gamma} dz}}\, ,
\end{equation}
where now $\Phi$ is the angle between $\kphp$ and $\pp$. Hence, we
evaluate these angular harmonics at fixed photon-jet transverse
momentum imbalance $Q_\perp$ and average transverse momentum
$\tilde{P}_\perp$ by averaging over their relative
orientation. Because of the requirement that
$Q_\perp\ll\tilde{P}_\perp$ the angle $\Phi$ is close to $\pm\pi$. Up
to order $(Q_\perp/\tilde{P}_\perp)^2$ we have
\begin{eqnarray}
\cos\Phi &=& -1 + \frac{1}{4} \left(\frac{Q_\perp}{\tilde{P}_\perp}\right)^2
\left(1-\cos 2\phi\right) ~, \\
\cos2\Phi &=& 1 - \left(\frac{Q_\perp}{\tilde{P}_\perp}\right)^2
\left(1-\cos 2\phi\right) ~,\\
\cos3\Phi &=& -1 + \frac{9}{4}\left(\frac{Q_\perp}{\tilde{P}_\perp}\right)^2
\left(1-\cos 2\phi\right) ~.
\end{eqnarray}
Since $\langle\cos 2\phi\rangle=a_2$ is itself of order
$(Q_\perp/\tilde{P}_\perp)^2$ we have, to this order, the following predictions:
\begin{eqnarray}
v_1^2\equiv \langle\cos\Phi\rangle &=& -1 + \frac{1}{4}
\left(\frac{Q_\perp}{\tilde{P}_\perp}\right)^2 = -1 +
\frac{|\kphp+\pp|^2}{|\kphp-\pp|^2} ~, \\
v_2^2\equiv \langle\cos2\Phi\rangle &=& 1 -
\left(\frac{Q_\perp}{\tilde{P}_\perp}\right)^2 = 1 - 4
\frac{|\kphp+\pp|^2}{|\kphp-\pp|^2}~,\\
v_3^2\equiv \langle\cos3\Phi\rangle &=& -1 + \frac{9}{4}
\left(\frac{Q_\perp}{\tilde{P}_\perp}\right)^2 = -1 + 9
\frac{|\kphp+\pp|^2}{|\kphp-\pp|^2}~.
\end{eqnarray}
These expressions provide the leading dependence of $v_1^2$, $v_2^2$,
$v_3^2$ on $Q_\perp/\tilde{P}_\perp$. These moments are insensitive to
the transverse momentum dependence of the gluon distributions $F_i$.

\section{Summary and discussion}
\label{sec:summ}

In this paper we have computed the photon-jet cross section
corresponding to the process $g\to q\gamma$ in the small-$x$ regime of
p+A collisions. We focus on nearly back to back
configurations with a photon-jet transverse momentum imbalance
$\Qp=\kphp+\pp$ of much smaller magnitude than their average
transverse momentum $\Ppt=(\pp-\kphp)/2$. In this limit the cross
section for the process can be expressed in terms of transverse
momentum dependent gluon distributions for the heavy-ion target. We
determine which gluon distributions appear in this process and find,
in particular, that the linearly polarized distributions do not enter
(similar to $q\to q\gamma$). We also show that the
contribution from the leading order $q\to q\gamma$ process is
suppressed by two powers of $Q_\perp/Q_S$ as compared to
$g\to q\gamma$ if $Q_\perp<Q_S$ is less than the saturation scale of
the nucleus.

The calculation begins by considering scattering of a gluon from the
projectile proton off the strong color field of the target ion,
thereby producing a quark, an anti-quark, and a photon. The $q\gamma$
final state is obtained by integrating over the quark from which the
photon was emitted (in the collinear approximation). We find that the
transverse momentum scale that appears in the gluon distributions of
the target is given by the left-over transverse momentum $1-z$ of the
quark relative to that of the photon, $z$, times the hard scale
$\tilde{P}_\perp$.
Hence, configurations
where the photon picks up most of the momentum of the parent quark,
$1-z\ll1$, so that a second hadronic jet collinear to the photon is
not observed (due to isolation cuts), do probe the gluon distributions
of the target in the non-linear regime.

Finally, we provide analytic expressions and qualitative numerical
estimates for $a_1=\langle\cos \phi\rangle$ and $a_2=\langle\cos
2\phi\rangle$ angular moments, where $\phi$ denotes the angle between
$\Qp$ and $\Ppt$. We predict that $a_2\ll a_1$ due to a power
suppression by a factor of $Q_\perp/\tilde{P}_\perp$. The
$\tilde{P}_\perp$-dependence of these angular correlations provides
insight into the transverse momentum dependence of the gluon
distributions of the target. In particular, numerically large $a_1$
and $a_2$ are obtained for more restrictive photon isolation cuts, and
when the transverse momentum scale in the gluon distributions is on
the order of a few times the saturation scale of the heavy ion target.

We have also derived analytic estimates
for the ``azimuthal angular harmonics'' $v^2_n=\langle\cos n\Phi\rangle$
up to $n=3$ and ${\cal O}(Q_\perp^2/\tilde{P}_\perp^2)$, where now
$\Phi$ denotes the angle between the transverse momentum $\kphp$ of
the photon and that of the jet, $\pp$. However, up to order
$(Q_\perp/\tilde{P}_\perp)^2$ these $v_n$ moments are insensitive to
the transverse momentum dependence of the gluon distributions.

If photon-jet angular correlations in p+A collisions can
indeed be studied experimentally at high-energy colliders then a more
quantitative evaluation of the angular distributions than presented in
this initial study would be warranted. Most importantly, small-$x$
evolution effects on the transverse momentum dependent gluon
distributions should be incorporated. This could be done by
substituting the solution of the BK equation for
$\Gamma_{x_A}(x_\perp^2)$ into the expressions for
$F_i(x_A,k_\perp^2)$ given in sec.~\ref{sec:F_H_distributions}.

One should also account for the Sudakov suppression which arises due to the
presence of the two scales $Q_\perp$ and
$\tilde{P}_\perp$~\cite{Boer:2017xpy,Sudakov}. On the other hand, for
$\tilde{P}_\perp$ only a few times greater than $Q_\perp$ (to suppress
power corrections reasonably well) $\log \tilde{P}_\perp/Q_\perp$
is numerically not much greater than 1 and we would not expect a very
strong effect on the angular distributions. In any case, the results
presented here for the $g\to q\gamma$ process could be used as a
starting point for such improvements.
Our present analysis already suggests that photon-jet
correlations in p+A should provide valuable insight
into transverse momentum dependent
gluon distributions in the regime of non-linear color fields.

\acknowledgments{
S.~B. was supported by the European Union Seventh Framework Programme (FP7 2007-2013) under
grant agreement No. 291823, Marie Curie FP7-PEOPLE-2011-COFUND NEWFELPRO Grant No. 48. S.~B. also acknowledges the support of HZZO Grant No. 8799.
A.~D.\ gratefully acknowledges support by the DOE Office of Nuclear
Physics through Grant No.\ DE-FG02-09ER41620; and from The City
University of New York through the PSC-CUNY Research grant 60262-0048.
}

\appendix

\section{Expansion in powers of gluon momenta}
\label{sec:expan}

Here we perform an expansion of the amplitude for $g \to q\qbar\gamma$, given in \eqref{eq:full-amp}, in powers of gluon momenta from the proton and from the nucleus.
We start by generalizing the expressions for the matrices $R_\beta^\mu$ from Eqs.~\eqref{eq:R1R2mu}-\eqref{eq:R12mu} as
\begin{equation}
\begin{split}
& R_1^{\mu\nu} \equiv -\frac{1}{P^2}\gamma^\mu \frac{\ws + m}{w^2 - m^2}\gamma^\nu\,,\\
& R_2^{\mu\nu} \equiv -\frac{1}{P^2}\gamma^\nu \frac{-\vs + m}{v^2 - m^2}\gamma^\mu\,,\\
& R_9^{\mu\nu\alpha_1\alpha_2}  \equiv (-i)\int_{-\infty}^\infty \frac{dk^-}{2\pi}  \tilde{R}_9^{\mu\nu\alpha_1\alpha_2}\,, \qquad \tilde{R}_9^{\mu\nu\alpha_1\alpha_2} \equiv\gamma^\mu\frac{\ws + m}{w^2 - m^2 + i\epsilon}\gamma^{\alpha_1} \frac{\us + m}{u^2 - m^2 + i\epsilon}\gamma^\nu \frac{\ls + m}{l^2 - m^2 + i\epsilon}\gamma^{\alpha_2}\\
& R_{10}^{\mu\nu\alpha_1\alpha_2}  \equiv (-i)\int_{-\infty}^\infty \frac{dk^-}{2\pi} \tilde{R}_{10}^{\mu\nu\alpha_1\alpha_2} \,, \qquad \tilde{R}_{10}^{\mu\nu\alpha_1\alpha_2} \equiv \gamma^{\alpha_1}\frac{\gs + m}{g^2 - m^2 + i\epsilon}\gamma^\nu \frac{\hs + m}{h^2 - m^2 + i\epsilon}\gamma^{\alpha_2} \frac{-\vs + m}{v^2 - m^2 + i\epsilon}\gamma^\mu \,,\\
& R_{11}^{\mu\nu\alpha_1\alpha_2}  \equiv (-i)\int_{-\infty}^\infty \frac{dk^-}{2\pi} \tilde{R}_{11}^{\mu\nu\alpha_1\alpha_2}\,, \qquad \tilde{R}_{11}^{\mu\nu\alpha_1\alpha_2} \equiv \gamma^{\alpha_1}\frac{\gs + m}{g^2 - m^2 + i\epsilon}\gamma^\mu \frac{\us + m}{u^2 - m^2 + i\epsilon}\gamma^{\nu} \frac{\ls + m}{l^2 - m^2 + i\epsilon}\gamma^{\alpha_2}\,,\\
& R_{12}^{\mu\nu\alpha_1\alpha_2}  \equiv (-i)\int_{-\infty}^\infty \frac{dk^-}{2\pi} \tilde{R}_{12}^{\mu\nu\alpha_1\alpha_2} \,,\qquad \tilde{R}_{12}^{\mu\nu\alpha_1\alpha_2} \equiv \gamma^{\alpha_1}\frac{\gs + m}{g^2 - m^2 + i\epsilon}\gamma^\nu \frac{\hs + m}{h^2 - m^2 + i\epsilon}\gamma^\mu \frac{\ls + m}{l^2 - m^2 + i\epsilon}\gamma^{\alpha_2}\,,\\
\end{split}
\end{equation}
and also
\begin{equation}
R_g^{\mu\nu} = \sum_{\beta = 1}^2 R_\beta^{\mu\nu} \,, \qquad R_{q\bar{q}}^{\mu\nu\alpha_1\alpha_2} = \sum_{\beta = 9}^{12} R_\beta^{\mu\nu\alpha_1\alpha_2}\,. 
\label{eq:RgRqqbar}
\end{equation}
Performing the $k^-$ integration in $R_{\beta}^{\mu\nu\alpha_1\alpha_2}$, $\beta = 9,\dots,12$, we can show that $T_{q\bar{q}}^{\mu} = R_{q\bar{q}}^{\mu - ++}$ and obviously $T_g^\mu = R_g^{\mu\nu}C_{L \nu} $.

Contracting $R_\beta^{\mu\nu++}$ for $\beta=9,\dots,12$ with $k_{1\nu}$ we get
\begin{equation}
\begin{split}
& k_{1\nu}R_9^{\mu\nu ++} = -\frac{\gamma^\mu(\ws + m)\gamma^+}{w^2 - m^2} = P^2 R_1^{\mu +} \,,\\
& k_{1\nu}R_{10}^{\mu\nu ++} = -\frac{\gamma^+(-\vs + m)\gamma^\mu}{v^2 - m^2} = P^2 R_2^{\mu +} \,,\\
& k_{1\nu}R_{11}^{\mu\nu ++} = -\frac{\gamma^+(\gs + m)\gamma^\mu(\ls + m)\gamma^+}{S} = -k_{1\nu}R_{12}^{\mu\nu ++} \,,\\
\end{split}
\end{equation}
Decomposing these expressions we can use $k_1^+  R_{q\bar{q}}^{\mu - ++} = P^+ R_{q\bar{q}}^{\mu - ++}$ to redefine $T_g^\mu$.
We can write the amplitude as
\begin{equation}
  \begin{split}
    \calM^\mu(\MDEP) &= -q_f e g^2\int_{\kp \khp}\int_{\xp\yp}
    \frac{\rhop^a(\khp)}{k_{1\perp}^2}\,
    e^{i\kp\cdot\xp+i(\kAp-\kp)\cdot\yp} \\
    &\qquad\times \ubar \bigl\{ R_g^{\mu\nu}C_\nu(P,\khp)U(\xp)^{ba}t^b
    -\frac{k_{1 i}}{P^+}R^{\mu i ++}_{q\bar{q}}\Uf(\xp)t^a\Uf^\dag(\yp)\bigr\}\vp\,,
  \end{split}
\label{eq:full-amp2}
\end{equation}
with
\begin{equation}
  C^+(q,\khp) = 0\,,\quad
  C^-(q,\khp)
    = \frac{-2\khp\cdot(\qp-\khp)}{q^+ +i\epsilon}\,,\quad
  \bC_\perp(q;\khp)
    = \frac{\qp \, k_{1\perp}^2}{(q^+ +i\epsilon)(q^- +i\epsilon)} - 2\khp\,.
\end{equation}

Contracting $\tilde{R}_\beta^{\mu\nu\alpha_1\alpha_2}$ for $\beta=9,\dots,12$ with $k_{\alpha_1}$ or with $(k_2 - k)_{\alpha_2}$ we get
\begin{equation}
\begin{split}
& k_{\alpha_1} \tilde{R}_9^{\mu\nu\alpha_1\alpha_2} = -\gamma^\mu \frac{\ws + m}{w^2 - m^2 + i\epsilon}\gamma^\nu \frac{\ls + m}{l^2 - m^2 + i\epsilon}\gamma^{\alpha_2}  + \gamma^\mu \frac{\us + m}{u^2 - m^2 + i\epsilon}\gamma^\nu \frac{\ls + m}{l^2 - m^2 + i\epsilon}\gamma^{\alpha_2}\,,\\
& k_{\alpha_1}\tilde{R}_{10}^{\mu\nu\alpha_1\alpha_2} =  -\gamma^\nu \frac{\hs + m}{h^2 - m^2 + i\epsilon} \gamma^{\alpha_2} \frac{-\vs + m}{v^2 - m^2 + i\epsilon}\gamma^\mu \,,\\
& k_{\alpha_1} \tilde{R}_{11}^{\mu\nu\alpha_1\alpha_2} = -\gamma^\mu \frac{\us + m}{u^2 - m^2 + i\epsilon}\gamma^\nu \frac{\ls + m}{l^2 - m^2 + i\epsilon}\gamma^{\alpha_2}\,,\\
& k_{\alpha_1} \tilde{R}_{12}^{\mu\nu\alpha_1\alpha_2} = -\gamma^\nu \frac{\hs + m}{h^2 - m^2+i\epsilon}\gamma^\mu \frac{\ls + m}{l^2 - m^2 + i\epsilon}\gamma^{\alpha_2}\,.\\
\end{split}
\label{eq:winuc1}
\end{equation} 

\begin{equation}
\begin{split}
& (k_2 - k)_{\alpha_2} \tilde{R}_9^{\mu\nu\alpha_1\alpha_2} = \gamma^\mu \frac{\ws + m}{w^2 - m^2 + i\epsilon}\gamma^{\alpha_1} \frac{\us + m}{u^2 - m^2 + i\epsilon}\gamma^\nu \,,\\
& (k_2 - k)_{\alpha_2}\tilde{R}_{10}^{\mu\nu\alpha_1\alpha_2} =  \gamma^{\alpha_1} \frac{\gs + m}{g^2 - m^2 + i\epsilon} \gamma^\nu \frac{-\vs + m}{v^2 - m^2 + i\epsilon}\gamma^\mu
- \gamma^{\alpha_1} \frac{\gs + m}{g^2 - m^2 + i\epsilon} \gamma^\nu \frac{\hs + m}{h^2 - m^2 + i\epsilon}\gamma^\mu \,,\\
& (k_2 - k)_{\alpha_2} \tilde{R}_{11}^{\mu\nu\alpha_1\alpha_2} = \gamma^{\alpha_1} \frac{\gs + m}{g^2 - m^2 + i\epsilon}\gamma^\mu \frac{\us + m}{u^2 - m^2 + i\epsilon}\gamma^\nu\,,\\
& (k_2 - k)_{\alpha_2} \tilde{R}_{12}^{\mu\nu\alpha_1\alpha_2} = \gamma^{\alpha_1} \frac{\gs + m}{g^2 - m^2+i\epsilon}\gamma^\nu \frac{\hs + m}{h^2 - m^2 + i\epsilon}\gamma^\mu\,.\\
\end{split}
\label{eq:winuc2}
\end{equation} 

Next, we use Eq.~\eqref{eq:winuc1} to combine $\tilde{R}_9^{\mu\nu\alpha_1\alpha_2}$ with $\tilde{R}_{11}^{\mu\nu\alpha_1\alpha_2}$
\begin{equation}
k^-(\tilde{R}_9^{\mu\nu ++} + \tilde{R}_{11}^{\mu\nu ++}) = -k_{i_1}(\tilde{R}_9^{\mu\nu i_1 +} + \tilde{R}_{11}^{\mu\nu i_1 +}) - \gamma^\mu \frac{\ws + m}{w^2 - m^2 + i\epsilon}\gamma^\nu \frac{\ls + m}{l^2 - m^2 + i\epsilon}\gamma^+\,,
\label{eq:ward3}
\end{equation} 
and we use \eqref{eq:winuc2} to combine $\tilde{R}_{10}^{\mu\nu\alpha_1\alpha_2}$ with $\tilde{R}_{12}^{\mu\nu\alpha_1\alpha_2}$
\begin{equation}
(k_2 - k)^-(\tilde{R}_{10}^{\mu\nu ++} + \tilde{R}_{12}^{\mu\nu ++}) = -(k_2 - k)_{i_2}(\tilde{R}_{10}^{\mu\nu + i_2} + \tilde{R}_{12}^{\mu\nu + i_2}) + \gamma^+ \frac{\gs + m}{g^2 - m^2 + i\epsilon} \gamma^\nu \frac{-\vs + m}{v^2 - m^2 + i\epsilon}\gamma^\mu\,.
\label{eq:ward4}
\end{equation} 
We first integrate these expressions over $k^-$. On the left hand side we get $R^{\mu\nu ++}_9 +R^{\mu\nu ++}_{11}$ and $R^{\mu\nu ++}_{10} + R^{\mu\nu ++}_{12}$, respectively.
Taking the limit $k_{1\perp}, k_{2\perp}, k_\perp \to 0$
we find that in Eq.~\eqref{eq:ward3} $k^- = P^-$, while in Eq.~\eqref{eq:ward4} $k^- = 0$. Combining these results together we get the following expression for $R^{\mu\nu++}_{q\qbar}$ 
\begin{equation}
\begin{split}
R^{\mu\nu ++}_{q\bar{q}} &= \gamma^\mu \frac{\ws + m}{w^2 - m^2} \frac{\kps}{P^-}\frac{\us + m}{u^2 - m^2}\gamma^\nu + \frac{\kps}{P^-}\frac{\gs + m}{g^2 - m^2}\gamma^\mu \frac{\us + m}{u^2 - m^2}\gamma^\nu + \frac{\kps}{P^-} \frac{\gs + m}{g^2 - m^2} \gamma^\nu \frac{-\vs + m}{v^2 - m^2}\gamma^\mu\\
&-\gamma^\nu \frac{\hs + m}{h^2 - m^2}\frac{\kAps - \kps}{P^-}\frac{-\vs + m}{v^2 - m^2}\gamma^\mu-\gamma^\nu \frac{\hs + m}{h^2 - m^2}\gamma^\mu \frac{\ls + m}{l^2 - m^2}\frac{\kAps - \kps}{P^-}-\gamma^\mu \frac{\ws + m}{w^2 - m^2}\gamma^\nu \frac{\ls + m}{l^2 - m^2}\frac{\kAps - \kps}{P^-}\\
& + \frac{1}{P^-}\gamma^\mu \frac{\ws + m}{w^2 - m^2}\gamma^\nu + \frac{1}{P^-}\gamma^\nu \frac{-\vs + m}{v^2 - m^2} \gamma^\mu \,,\\
\end{split}
\label{eq:Rqqb}
\end{equation}
where the first line comes from Eq.~\eqref{eq:ward3} and so we should understand $k^- = P^-$, while the second line comes from Eq.~\eqref{eq:ward4} and so $k^- = 0$. The third line can be joined to $R^{\mu\nu}_g$ (defined in Eq.~\eqref{eq:RgRqqbar}) which leads us to the final expression for the amplitude given in Eqs.~\eqref{eq:amp3}-\eqref{eq:har1}.

\section{Distribution functions} \label{sec:F_H_distributions}

We can write more explicit expressions for the distribution functions
$F_i$ and $H_i$ by using the large $N_c$ expression for the four
Wilson line correlator $C(x_A,\xp,\yp,\yp',\xp')$ in a Gaussian
model~\cite{Blaizot:2004wv,Fukushima:2007dy}
\begin{equation}
C(x_A,\xp,\yp,\yp',\xp') = \frac{N_c^2}{2} S((\xp - \xp')^2) S((\yp - \yp')^2)\,,
\label{eq:ClargeN}
\end{equation}
where $S(x_\perp^2) = e^{-\Gamma(x_\perp^2)}$ is the $S$-matrix for a
fundamental dipole of size $x_\perp$.
Within the McLerran-Venugopalan (MV) model \cite{MV} (without small-$x$ evolution) we have
\begin{equation}
\Gamma(x_\perp^2) = Q_S^2 \int_{\yp}[G_0(\yp) - G_0(\yp - \xp)]^2\,, \qquad Q_S^2 \equiv \frac{g^4 (N_c^2-1)}{4\pi N_c} \mu_A^2\,,
\end{equation}
$\mu_A^2$ is the conventional MV model parameter, i.e.\ the average valence
color charge density squared per unit transverse area, $Q_S^2$ is the saturation scale, and $G_0(\xp)$
is a solution of $\partial_\perp^2 G_0(\xp) =
\delta^{(2)}(\xp)$. Regularizing the IR divergence in the Fourier
transform of $G_0(\xp)$ as $G_0(\kp) = (\kp^2 + \Lambda^2)^{-1}$ we
can find
\begin{equation}
\Gamma(x_\perp^2) = \frac{Q_S^2}{2 \Lambda^2}\left[1-(x_\perp \Lambda) K_1(x_\perp\Lambda)\right]\,,
\end{equation}
where $K_1(x)$ is the modified Bessel function of the second kind.
Expanding around $x_\perp\Lambda \to 0$ we find
\begin{equation}
\Gamma(x_\perp^2) = \frac{x_\perp^2 Q_S^2}{8} \left(1-2\gamma_E + \log 4 - \log\frac{1}{x_\perp^2 \Lambda^2}\right) = \frac{x_\perp^2 Q_S^2}{8} \log\left(\frac{1}{x_\perp^2 \LIR^2}\right)\,, \qquad \LIR^2 \equiv \frac{\Lambda^2}{4e^{1-2\gamma_E}}\,.
\end{equation}

However, Eq.~\eqref{eq:ClargeN} is not restricted to the MV model. One
can include small-$x$ evolution effects by solving the BK
equation~\cite{BK} for $\Gamma(x_\perp^2)$ which now becomes also a
function of $x_A$: $\Gamma(x_\perp^2)\to\Gamma_{x_A}(x_\perp^2)$. We
proceed with this more general formulation and express the
distribution functions in terms of $\Gamma_{x_A}(x_\perp^2)$. We find
\begin{equation}
\begin{split}
&F_1(x_A,k_{2\perp}^2) = (\pi R_A^2) \, 4\pi N_c^2\int_0^\infty x_\perp d x_\perp J_0(k_{2\perp}x_\perp)\left\{\Gamma_{x_A}^{(1)}(x_\perp^2)+x_\perp^2 \left[\Gamma_{x_A}^{(2)}(x_\perp^2) - (\Gamma_{x_A}^{(1)}(x_\perp^2))^2\right]\right\}e^{-2\Gamma_{x_A}(x_\perp^2)}\,,\\
&F_2(x_A,k_{2\perp}^2) = -(\pi R_A^2) \, 4\pi N_c^2\int_0^\infty x_\perp d x_\perp J_0(k_{2\perp}x_\perp)x_\perp^2  (\Gamma_{x_A}^{(1)}(x_\perp^2))^2 e^{-2\Gamma_{x_A}(x_\perp^2)}\,,\\
&F_3(x_A,k_{2\perp}^2) = (\pi R_A^2) \, 4\pi N_c^2\int_0^\infty x_\perp d x_\perp J_0(k_{2\perp}x_\perp)\left\{\Gamma_{x_A}^{(1)}(x_\perp^2)+x_\perp^2 \left[\Gamma_{x_A}^{(2)}(x_\perp^2) - 2(\Gamma_{x_A}^{(1)}(x_\perp^2))^2\right]\right\}e^{-2\Gamma_{x_A}(x_\perp^2)}\,,\\
\end{split}
\end{equation}
\begin{equation}
\begin{split}
&H_1(x_A,k_{2\perp}^2) = -(\pi R_A^2) \, 4\pi N_c^2\int_0^\infty x_\perp d x_\perp J_2(k_{2\perp}x_\perp)x_\perp^2 \left[\Gamma_{x_A}^{(2)}(x_\perp^2) - (\Gamma_{x_A}^{(1)}(x_\perp^2))^2\right]e^{-2\Gamma_{x_A}(x_\perp^2)}\,,\\
&H_2(x_A,k_{2\perp}^2) = (\pi R_A^2) \, 4\pi N_c^2\int_0^\infty x_\perp d x_\perp J_2(k_{2\perp}x_\perp)x_\perp^2  (\Gamma_{x_A}^{(1)}(x_\perp^2))^2 e^{-2\Gamma_{x_A}(x_\perp^2)}\,,\\
&H_3(x_A,k_{2\perp}^2) = -(\pi R_A^2) \, 4\pi N_c^2\int_0^\infty x_\perp d x_\perp J_2(k_{2\perp}x_\perp)x_\perp^2 \left[\Gamma_{x_A}^{(2)}(x_\perp^2) - 2(\Gamma_{x_A}^{(1)}(x_\perp^2))^2\right]e^{-2\Gamma_{x_A}(x_\perp^2)}\,,\\
\end{split}
\label{eq:His}
\end{equation}
where $\Gamma_{x_A}^{(n)}(x_\perp^2) \equiv d^n\Gamma_{x_A}(x_\perp^2)/d(x_\perp^2)^n$.
In the MV model the $F$-functions become
\begin{equation}
\begin{split}
&F_1(x_A,k_{2\perp}^2) = (\pi R_A^2) \frac{Q_S^2 N_c^2}{16}\int_0^\infty x_\perp d x_\perp J_0(k_{2\perp}x_\perp)\left(8\pi \log\frac{1}{x_\perp^2 \LIR^2} - \pi x_\perp^2 Q_S^2\log^2\frac{1}{x_\perp^2\LIR^2}\right)e^{-\frac{Q_S^2 x_\perp^2}{4}\log\frac{1}{x_\perp^2 \LIR^2}}\,,\\
&F_2(x_A,k_{2\perp}^2) = (\pi R_A^2) \frac{\pi Q_S^2 N_c^2}{8}\int_0^\infty x_\perp d x_\perp J_0(k_{2\perp}x_\perp)x_\perp^2 Q_S^2 \log^2\frac{1}{x_\perp^2\LIR^2}e^{-\frac{Q_S^2 x_\perp^2}{4}\log\frac{1}{x_\perp^2 \LIR^2}}\,,\\
&F_3(x_A,k_{2\perp}^2) = (\pi R_A^2) \frac{Q_S^2 N_c^2}{8}\int_0^\infty x_\perp d x_\perp J_0(k_{2\perp}x_\perp)\left(4\pi \log\frac{1}{x_\perp^2 \LIR^2} - \pi Q_S^2 x_\perp^2\log^2\frac{1}{x_\perp^2\LIR^2}\right)e^{-\frac{Q_S^2 x_\perp^2}{4}\log\frac{1}{x_\perp^2 \LIR^2}}\,.
\end{split}
\end{equation}

\end{document}